\documentclass[aps,twocolumn,prc,floatfix,preprintnumbers,superscriptaddress,notitlepage,nofootinbib]{revtex4-1}

\usepackage[utf8]{inputenc}
\usepackage{graphicx}% Include figure files
\usepackage{dcolumn}% Align table columns on decimal point
\usepackage{bm}% bold math
\usepackage{lineno}
\usepackage{float}
\usepackage{multirow}
\usepackage{xspace}
\usepackage[section]{placeins}
\usepackage{xcolor}
\usepackage{amsmath}
\usepackage{gensymb}
\usepackage[colorlinks = true,
            linkcolor = blue,
            urlcolor  = blue,
            citecolor = blue,
            anchorcolor = blue]{hyperref}
\begin{document}

\newcommand{\sigreac}{$\sigma_{reac}$\xspace}
\newcommand\czer[1]{{\color{red}#1}}
%\linenumbers

\newcommand*{\PITT}{University of Pittsburgh, Department of Physics and Astronomy, Pittsburgh, PA 15260, USA}
\newcommand*{\PITTIndex}{1}
\affiliation{\PITT}
\newcommand*{\FNAL}{Fermi National Accelerator Laboratory, Batavia, IL, USA} 
\newcommand*{\FNALIndex}{2}
\affiliation{\FNAL}
\newcommand*{\ODU}{Old Dominion University, Norfolk, Virginia 23529}
\newcommand*{\ODUindex}{3}
\affiliation{\ODU}
\newcommand*{\TAU}{Tel Aviv University, Tel Aviv-Yafo 6997801, Israel}
\newcommand*{\TAUindex}{4}
\affiliation{\TAU}
\newcommand*{\LANL}{Los Alamos National Laboratory (LANL), Los Alamos, NM, 87545, USA}
\newcommand*{\LANLindex}{5}
\affiliation{\LANL}
\newcommand*{\ANL}{Argonne National Laboratory, Argonne, Illinois 60439}
\newcommand*{\ANLindex}{6}
\affiliation{\ANL}
\newcommand*{\ASU}{Arizona State University, Tempe, Arizona 85287-1504}
\newcommand*{\ASUindex}{7}
\affiliation{\ASU}
\newcommand*{\CSUDH}{California State University, Dominguez Hills, Carson, CA 90747}
\newcommand*{\CSUDHindex}{8}
\affiliation{\CSUDH}
\newcommand*{\CANISIUS}{Canisius University, Buffalo, NY}
\newcommand*{\CANISIUSindex}{9}
\affiliation{\CANISIUS}
\newcommand*{\CMU}{Carnegie Mellon University, Pittsburgh, Pennsylvania 15213}
\newcommand*{\CMUindex}{10}
\affiliation{\CMU}
\newcommand*{\SACLAY}{IRFU, CEA, Universit'{e} Paris-Saclay, F-91191 Gif-sur-Yvette, France}
\newcommand*{\SACLAYindex}{11}
\affiliation{\SACLAY}
\newcommand*{\CNU}{Christopher Newport University, Newport News, Virginia 23606}
\newcommand*{\CNUindex}{12}
\affiliation{\CNU}
\newcommand*{\UCONN}{University of Connecticut, Storrs, Connecticut 06269}
\newcommand*{\UCONNindex}{13}
\affiliation{\UCONN}
\newcommand*{\DUQUESNE}{Duquesne University, 600 Forbes Avenue, Pittsburgh, PA 15282 }
\newcommand*{\DUQUESNEindex}{14}
\affiliation{\DUQUESNE}
\newcommand*{\FU}{Fairfield University, Fairfield CT 06824}
\newcommand*{\FUindex}{15}
\affiliation{\FU}
\newcommand*{\FERRARAU}{Universita' di Ferrara , 44121 Ferrara, Italy}
\newcommand*{\FERRARAUindex}{16}
\affiliation{\FERRARAU}
\newcommand*{\FIU}{Florida International University, Miami, Florida 33199}
\newcommand*{\FIUindex}{17}
\affiliation{\FIU}
\newcommand*{\GWUI}{The George Washington University, Washington, DC 20052}
\newcommand*{\GWUIindex}{18}
\affiliation{\GWUI}
\newcommand*{\GSIFFN}{GSI Helmholtzzentrum fur Schwerionenforschung GmbH, D-64291 Darmstadt, Germany}
\newcommand*{\GSIFFNindex}{19}
\affiliation{\GSIFFN}
\newcommand*{\INFNFE}{INFN, Sezione di Ferrara, 44100 Ferrara, Italy}
\newcommand*{\INFNFEindex}{20}
\affiliation{\INFNFE}
\newcommand*{\INFNFR}{INFN, Laboratori Nazionali di Frascati, 00044 Frascati, Italy}
\newcommand*{\INFNFRindex}{21}
\affiliation{\INFNFR}
\newcommand*{\INFNGE}{INFN, Sezione di Genova, 16146 Genova, Italy}
\newcommand*{\INFNGEindex}{22}
\affiliation{\INFNGE}
\newcommand*{\INFNRO}{INFN, Sezione di Roma Tor Vergata, 00133 Rome, Italy}
\newcommand*{\INFNROindex}{23}
\affiliation{\INFNRO}
\newcommand*{\INFNTUR}{INFN, Sezione di Torino, 10125 Torino, Italy}
\newcommand*{\INFNTURindex}{24}
\affiliation{\INFNTUR}
\newcommand*{\INFNCAT}{INFN, Sezione di Catania, 95123 Catania, Italy}
\newcommand*{\INFNCATindex}{25}
\affiliation{\INFNCAT}
\newcommand*{\INFNPAV}{INFN, Sezione di Pavia, 27100 Pavia, Italy}
\newcommand*{\INFNPAVindex}{26}
\affiliation{\INFNPAV}
\newcommand*{\ORSAY}{Universit'{e} Paris-Saclay, CNRS/IN2P3, IJCLab, 91405 Orsay, France}
\newcommand*{\ORSAYindex}{27}
\affiliation{\ORSAY}
\newcommand*{\KSU}{King Saud University, Riyadh, Kingdom of Saudi Arabia}
\newcommand*{\KSUindex}{28}
\affiliation{\KSU}
\newcommand*{\KNU}{Kyungpook National University, Daegu 41566, Republic of Korea}
\newcommand*{\KNUindex}{29}
\affiliation{\KNU}
\newcommand*{\LAMAR}{Lamar University, 4400 MLK Blvd, PO Box 10046, Beaumont, Texas 77710}
\newcommand*{\LAMARindex}{30}
\affiliation{\LAMAR}
\newcommand*{\MIT}{Massachusetts Institute of Technology, Cambridge, Massachusetts  02139-4307}
\newcommand*{\MITindex}{31}
\affiliation{\MIT}
\newcommand*{\MISS}{Mississippi State University, Mississippi State, MS 39762-5167}
\newcommand*{\MISSindex}{32}
\affiliation{\MISS}
\newcommand*{\UNH}{University of New Hampshire, Durham, New Hampshire 03824-3568}
\newcommand*{\UNHindex}{33}
\affiliation{\UNH}
\newcommand*{\NSU}{Norfolk State University, Norfolk, Virginia 23504}
\newcommand*{\NSUindex}{34}
\affiliation{\NSU}
\newcommand*{\OHIOU}{Ohio University, Athens, Ohio  45701}
\newcommand*{\OHIOUindex}{35}
\affiliation{\OHIOU}
\newcommand*{\JLUGiessen}{II Physikalisches Institut der Universitaet Giessen, 35392 Giessen, Germany}
\newcommand*{\JLUGiessenindex}{36}
\affiliation{\JLUGiessen}
\newcommand*{\RPI}{Rensselaer Polytechnic Institute, Troy, New York 12180-3590}
\newcommand*{\RPIindex}{37}
\affiliation{\RPI}
\newcommand*{\URICH}{University of Richmond, Richmond, Virginia 23173}
\newcommand*{\URICHindex}{38}
\affiliation{\URICH}
\newcommand*{\ROMAII}{Universita' di Roma Tor Vergata, 00133 Rome Italy}
\newcommand*{\ROMAIIindex}{39}
\affiliation{\ROMAII}
\newcommand*{\MSU}{Skobeltsyn Institute of Nuclear Physics, Lomonosov Moscow State University, 119234 Moscow, Russia}
\newcommand*{\MSUindex}{40}
\affiliation{\MSU}
\newcommand*{\SCAROLINA}{University of South Carolina, Columbia, South Carolina 29208}
\newcommand*{\SCAROLINAindex}{41}
\affiliation{\SCAROLINA}
\newcommand*{\TEMPLE}{Temple University,  Philadelphia, PA 19122 }
\newcommand*{\TEMPLEindex}{42}
\affiliation{\TEMPLE}
\newcommand*{\JLAB}{Thomas Jefferson National Accelerator Facility, Newport News, Virginia 23606}
\newcommand*{\JLABindex}{43}
\affiliation{\JLAB}
\newcommand*{\ULS}{Universidad de La Serena, Avda. Juan Cisternas 1200, La Serena, Chile}
\newcommand*{\ULSindex}{44}
\affiliation{\ULS}
\newcommand*{\UTFSM}{Universidad T\'{e}cnica Federico Santa Mar\'{i}a, Casilla 110-V Valpara\'{i}so, Chile}
\newcommand*{\UTFSMindex}{45}
\affiliation{\UTFSM}
\newcommand*{\BRESCIA}{Universit`{a} degli Studi di Brescia, 25123 Brescia, Italy}
\newcommand*{\BRESCIAindex}{46}
\affiliation{\BRESCIA}
\newcommand*{\UCR}{University of California Riverside, 900 University Avenue, Riverside, CA 92521, USA}
\newcommand*{\UCRindex}{47}
\affiliation{\UCR}
\newcommand*{\GLASGOW}{University of Glasgow, Glasgow G12 8QQ, United Kingdom}
\newcommand*{\GLASGOWindex}{48}
\affiliation{\GLASGOW}
\newcommand*{\UTK}{University of Tennessee, Knoxville, Tennessee 37996, USA}
\newcommand*{\UTKindex}{49}
\affiliation{\UTK}
\newcommand*{\YORK}{University of York, York YO10 5DD, United Kingdom}
\newcommand*{\YORKindex}{50}
\affiliation{\YORK}
\newcommand*{\VIRGINIA}{University of Virginia, Charlottesville, Virginia 22901}
\newcommand*{\VIRGINIAindex}{51}
\affiliation{\VIRGINIA}
\newcommand*{\WM}{College of William and Mary, Williamsburg, Virginia 23187-8795}
\newcommand*{\WMindex}{52}
\affiliation{\WM}
\newcommand*{\YEREVAN}{Yerevan Physics Institute, 375036 Yerevan, Armenia}
\newcommand*{\YEREVANindex}{53}
\affiliation{\YEREVAN}

\newcommand*{\NOWJLAB}{Thomas Jefferson National Accelerator Facility, Newport News, Virginia 23606}
\newcommand*{\NOWISU}{Idaho State University, Pocatello, Idaho 83209}
\newcommand*{\NOWCUA}{Catholic University of America, Washington, D.C. 20064}
\newcommand*{\NOWWASHU}{Washington University at Saint Louis, St. Louis, MO 63130}
 %%%%%%%%%%%%%%% END OF Latex Macros for institute addresses  %%%%%%%%%%%%%%%%%%%%%%%%% 

%\title{New methods and Results for Proton Transparency in Nuclei}
\title{Proton Transparency and Neutrino Physics: New Methods and Modeling}
\author{S.~Dytman}\affiliation{\PITT}
\author{M.~Betancourt}\affiliation{\FNAL}
\author{N.~Steinberg}\affiliation{\FNAL}
\author{L.B.~Weinstein}\affiliation{\ODU}
\author{A.~Ashkenazi}\affiliation{\TAU}
\author{J.~Tena-Vidal}\affiliation{\TAU}
\author{A.~Papadopoulou}\affiliation{\LANL}
\author{G.~Chambers-Wall} \affiliation{\FNAL}\altaffiliation[Current address:]{\NOWWASHU}
\author{J.~Smith}\affiliation{\FNAL}
\author {P.~Achenbach} 
\affiliation{\JLAB}
% ...

\collaboration{The CLAS Collaboration}
\noaffiliation

\date{\today}
%\date{\today}% It is always \today, today,
             %  but any date may be explicitly specified

\begin{abstract}
%abstract has to be short

Extracting accurate results from neutrino oscillation and cross section experiments requires accurate simulation of the neutrino-nucleus interaction.
The rescattering of outgoing hadrons (final state interactions) by the rest of the nucleus is an important component of these interactions. 
We present a new measurement of proton transparency (defined as the fraction of outgoing protons that emerge without significant rescattering) using electron-nucleus scattering data recorded by the CLAS detector at Jefferson Laboratory on helium, carbon, and iron targets. This analysis by the Electrons for Neutrinos ($e4\nu$) collaboration uses a new data-driven method to extract the transparency.  It defines transparency as the ratio of electron-scattering events with a detected proton to quasi-elastic electron-scattering events where a proton should have been knocked out. Our results are consistent with previous measurements that determined the transparency from the ratio of measured events to theoretically predicted events. 
We find that the GENIE event generator, which is widely used by oscillation experiments to simulate neutrino-nucleus interactions, needs to better describe both the nuclear ground state and proton rescattering in order to reproduce our measured transparency ratios, especially at lower proton momenta.
 
\end{abstract}

\maketitle

\section{Introduction}\label{introduction}

Accurate neutrino cross section measurements~\cite{nustec-review} and modeling of nuclear effects are required for precise measurements of neutrino oscillation results such as charge-parity (CP) symmetry violation and the ordering of the neutrino masses~\cite{MicroBooNE:2022sdp,NOvA:2021nfi,T2K:2023smv,DUNE:2022aul}. These experiments rely heavily on event generator codes~\cite{campbell2022}, simulations of neutrino-nucleus interactions used to estimate efficiencies, backgrounds, and systematic uncertainties for published results. 
%Trying to estimate sensitivity to these effects, neutrino oscillation measurements often have significant components of their systematic uncertainty due to FSI.  Although agreement with neutrino cross section data is generally qualitative, large disagreements are seldom easily attributable to FSI effects.

Event generator codes commonly used in neutrino experiments~\cite{Andreopoulos:2009rq,Golan:2012wx,hayato:2021heg} characterize experimental data according to basic processes through which the neutrino interacts with nucleons in the nuclear medium.  These processes include quasielastic (QE) scattering (single nucleon knockout due to one particle-one hole nuclear excitations), processes where the momentum transfer is shared by two nucleons (two-particle-two hole nuclear excitations ($2p2h$, sometimes called meson exchange currents (MEC))), and meson production (either through excitation of nucleon resonances (RES) or non-resonant (DIS) processes). The DIS processes include several interactions, ranging from single pion production at low energies to interactions with quarks at high energies\footnote{Existing neutrino event generators use a much broader definition of ``DIS" than is typical in hadronic physics.}.

The signature of this initial interaction (QE, MEC, RES or DIS) is often masked by final state interactions (FSI) where the outgoing hadrons interact in various ways with the residual nucleus.  For example, two nucleon knockout could be due to either a $2p2h$ interaction or to a QE interaction followed by rescattering of the struck nucleon. As a result, FSI models have a significant effect on background and efficiency estimation. FSI models are complicated because they involve the strong interaction at non-perturbative kinematics.  Therefore, there are always approximations and uncertainties in FSI models which can contribute significantly to the systematic uncertainties in many results.

Most event generator codes~\cite{GENIE:2021npt,Golan:2012wx,hayato:2021heg} validated FSI models against hadron-nucleus scattering data such as the total reaction cross section ($\sigma_{reac}$)~\cite{PinzonGuerra:2018rju,Carlson:1996ofz} which measures all inelastic processes.  
However, in neutrino-nucleus collisions, the outgoing hadrons start inside the nucleus and only traverse part of the nucleus as they exit, as opposed to hadron-nucleus scattering where the incident hadron starts outside the nucleus and potentially traverses the entire nucleus. In reality, hadron-nucleus cross sections can be dominated by collisions at the nuclear surface because the proton-nucleon interaction is very strong.  
Therefore, validation of FSI models for neutrino and electron scattering can be better accomplished against measurements of outgoing-hadron rescattering in lepton-nucleus collisions. To that end, electron scattering transparency experiments~\cite{Dutta:2012ii} measure the fraction of initial hadrons that emerge from the nucleus without significant rescattering.  

Neutrino- and electron-nucleus interactions are similar due to their shared origin in electroweak theory. Electrons interact via a vector current and  neutrinos interact via vector and axial-vector currents.  The nuclear structure and final state interactions of the outgoing hadrons are expected to be identical for both leptons except for details such as the lepton mass.  
These similarities are exploited by the Electrons for Neutrinos ($e4\nu$) project to use electron-nucleus scattering data to constrain neutrino-nucleus event generators~\cite{e4nu-nature}. 
Hadron transparency measurements are much easier with electrons than with neutrinos, since electron beams are mono-energetic and have much larger cross sections. All existing transparency data~\cite{Dutta:2003yt,Garrow:2001di,ONeill:1994znv,Garino:1992ca,HallC:2022qlb} have been determined with electron beams.

%Transparency measurements in the past mainly looked for deviations from standard nuclear theory that would be a signal for color transparency~\cite{Dutta:2012ii,CLAS:2012tlh}. Various measurements for protons~\cite{Dutta:2003yt,ONeill:1994znv,Garrow:2001di} and pions~\cite{Qian:2009aa} have been published for many nuclei. Additional measurements are available looking at nuclear structure~\cite{Rohe:2005vc,CLAS:2018xsb} and optical models~\cite{Garino:1992ca}. A previous CLAS analysis~\cite{CLAS:2018xsb} examined the effect for nucleon-nucleon (NN) correlations through ratios of proton transparency in nuclei to the value for a deuterium target.

Previous proton transparency measurements aimed both to understand the phenomenon itself and to search for possible deviations from standard nuclear theory that would be a signal for color transparency.  They compared the number of detected $(e,e'p)$ events to the number of expected $(e,e'p)$ events using different techniques.  

The most common method used quasielastic scattering at $x_B =Q^2/2m\nu\approx 1$ (where $Q^2$ is the squared four-momentum transfer, $m$ is the nucleon mass, and $\nu$ is the energy transfer~\cite{Garrow:2001di,Dutta:2003yt}). These events will be called ``true QE'' events in this article. They measured the number of $A(e,e'p)$ events with missing energy $E_{miss}\le 80$~MeV and missing momentum $p_{miss}\le 300$ MeV/c.  Here, $E_{miss}=\nu - T_p - T_{A-1}$ and $\vert\vec p_{miss}\vert = \vert \vec q - \vec p_p\vert$, where $T_p$ and $\vec p_p$ are the kinetic energy and momentum of the outgoing proton and $T_{A-1}=p_{miss}^2/2M_{A-1}$ is the kinetic energy of the residual nucleus. Then, the measured number of events was compared to a cross section calculation using a Plane Wave Impulse Approximation (PWIA):
\begin{equation}
  \sigma_{(e,e'p)} = K \sigma_{ep} S(p_i,E_{sep}),
\end{equation}
where $K$ is a kinematic factor, $\sigma_{ep}$ is the electron-offshell-proton cross section, and $S(p_i,E_{sep})$ is the spectral function which describes the probability of finding a proton in the nucleus with separation energy $E_{sep}$ and momentum $p_i$.  In the absence of final state interactions, $E_{miss} = E_{sep}$ and $p_{miss}=p_i$. The transparency is then:
\begin{equation}
    T_A^{previous}=\frac{N_{(e,e'p)}}{N^{PWIA}_{(e,e'p)}}.
    \label{TAold}
\end{equation}
Garino et al \cite{Garino:1992ca} defined the 
transparency slightly differently, namely as the double ratio of measured $(e,e'p)$ to $(e,e')$ yields for data and for a PWIA calculation. 

However, there are significant uncertainties in this method due to our poor knowledge of the spectral function.  Short range correlations~\cite{Hen:2016kwk} and other effects can reduce the spectral function probabilities at $E_{miss}\le 80$ MeV and  $p_{miss}\le 300$ MeV/c.  Most experiments~\cite{Garrow:2001di,Dutta:2003yt,ONeill:1994znv} used an Independent Particle Shell Model spectral function, and applied a correction factor ($1.11\pm 0.03$ for carbon and $1.26\pm 0.08$ for iron~\cite{Dutta:2003yt}) to account for the fraction of protons in short-range correlated (SRC) pairs with $p_{miss}$ and $E_{miss}$ outside of the experimental cuts \cite{ONeill:1994znv}.  Frankfurt, Strikman, and Zhalov~\cite{Frankfurt:2000ty} pointed out that these SRC correction factors greatly overestimate the SRC strength at $x_B  \approx 1$ where the parallel component of the initial nucleon momentum is small.  Neglecting these SRC correction factors would reduce the measured transparencies significantly. 

Other measurements support Frankfurt, Strikman, and Zhalov's argument.  Rohe et al \cite{Rohe:2005vc} compared their experimental yields to both PWIA and Correlated Basis Function (CBF) calculations.  They found $\approx 8\%$ larger transparencies with the CBF calculation.  Hen et al \cite{CLAS:2012usg} used a different technique, focusing on protons knocked out from SRC pairs at high missing momentum $300 \le p_{miss}\le 600$ MeV/c and $x_B\ge 1.2$.  The transparency was calculated as a double ratio. The numerator was the ratio of the number of detected $(e,e'p)$ events in nucleus $A$ to carbon and the denominator was the measured $(e,e')$ SRC ratio of nucleus $A$ to carbon~\cite{CLAS:2012usg,Hen:2016kwk}.  This method is independent of the effect of the SRC correction and the resulting transparencies are consistent with the small-SRC correction-factor of Ref.~\cite{Frankfurt:2000ty}.

Some comparisons of transparency data with event generators have been performed in recent years. Niewczas and Sobczyk~\cite{Niewczas:2019fro} compare the NuWro event generator~\cite{Golan:2012wx} to existing proton transparency data by applying the same acceptance criteria as was done in the experiments. Isaacson et al.~\cite{Isaacson:2020wlx} and Dytman et al.~\cite{Dytman:2021ohr} studied the theoretical inputs, showing the sensitivity to various elements of the models, especially the nuclear model. All these studies found approximate agreement with existing data. Although there is not a large discrepancy, sensitivity to both nuclear structure and FSI models was found for this comparison. All three articles call for improved precision and kinematic coverage in hadron transparency data to better test models.

To that end, we reanalyzed existing Jefferson Lab CLAS~\cite{mecking2003cebaf} electron-scattering data  to study proton transparency in a variety of nuclei to test neutrino event generator FSI models. 
Here, a different, more data-driven technique was used to extract the proton transparency.  It is defined as the ratio of the number of detected $A(e,e'p)$ events to the number of detected $A(e,e')$ events  that should have had an associated proton:  
\begin{equation}
    T_A^{new}=\frac{N_{(e,e'p)}}{N^{QE}_{(e,e')}}.
    \label{TAnew}
\end{equation}
As in previous measurements, the goal is to select only true QE events involving virtual photon absorption on a single proton. Selection of events for the numerator is similar to what was done in the past. However, unlike previous measurements, the denominator comes from inclusive data and all events in the numerator are a subset of these inclusive events. To ensure that the inclusive data are predominantly QE, we selected events on the low-energy-loss side of the QE peak to minimize the contribution of pion production and meson exchange current ($2p2h$) interactions.  Remaining background events were corrected for using the GENIE event generator~\cite{Andreopoulos:2009rq,e4nu-egenie:2020tbf}. The measured number of $(e,e')$ events was corrected downward for the fraction of neutron knockout events using the known electron-proton and electron-neutron elementary cross sections.  The number of $(e,e')$ events was corrected downward again to exclude the fraction of electrons scattering from nucleons in SRC pairs that would not be included in the $(e,e'p)$ sample. This SRC correction was calculated using a modern spectral function calculation~\cite{Betancourt:2023uxz,Benhar:1994hw}. 

We compared the results to the GENIE event generator.  GENIE simulates both neutrino-~\cite{Andreopoulos:2009rq,GENIE:2021npt} and electron-nucleus~\cite{e4nu-egenie:2020tbf} interactions. GENIE models agree qualitatively with $eA$ data in the regions dominated by QE scattering, which are relevant for this measurement. Disagreements between GENIE and electron-nucleon and electron-nucleus data in regions dominated by nucleon resonance excitation were seen. However, these regions are easily removed from the data. The new proton transparency measurements can be used to tune the GENIE simulation. This paper describes the experiment (Sect.~\ref{sec:methods}), analysis methods (Sect.~\ref{sec:analysis}), results (Sect.~\ref{sec:results}), discussion (Sect.~\ref{sec:discussion}), and conclusions (Sect~\ref{sec:conclusions}). 

\section{Experiment and Simulation}\label{sec:methods}
\noindent

The measurement reported here used data from the e2a run period of the Jefferson Lab CLAS spectrometer~\cite{mecking2003cebaf} at 2.261 and 4.461 GeV beam energies with helium, carbon, and iron targets. All targets had natural isotopic abundance, $^4$He (100\%), $^{12}$C (98.9\%) and $^{56}$Fe (91.8\%).  The first $e4\nu$ measurement~\cite{e4nu-nature} was based on these same data and many of the analysis techniques are reused here. 

CLAS had six separate but almost identical sectors with a total geometric acceptance much larger than the spectrometers previously used in transparency measurements. The sectors were arranged in the azimuthal direction around the beam line.  They are numbered clockwise as you look along the beam line away from the source with sector 1 on the left side looking downstream.  Each sector had three sets of drift chambers that tracked charged particles through the toroidal magnetic field. Following the drift chambers, each sector had a gas Cerenkov counter and an electromagnetic calorimeter for electron identification and triggering, and a scintillator time-of-flight array for timing and hadron identification.  The hadron $\phi$ acceptance ranged from $\approx 50\%$ at small angles to $\approx 85\%$ at $\theta\approx 90^\circ$.  
Differences among sectors can arise due to small shifts in magnetic coils and detectors and due to dead detector channels.  Yield differences among the sectors were included in the systematic uncertainties.

Monte Carlo (MC) simulations are important for establishing cuts, assessing background, and determining correction factors for this measurement. Two model sets of version 3.2.0 of the GENIE event generator~\cite{GENIE:2021npt,e4nu-egenie:2020tbf} were used. 

The G18 (specifically G18\_10a\_00\_00) model~\cite{e4nu-egenie:2020tbf} was the first implementation of electron scattering in GENIE. It describes QE scattering using the Rosenbluth cross section model with a Local Fermi Gas (LFG) model for the nucleon momentum with a fixed shift for binding energy. An empirical model~\cite{Katori:2013eoa} was used for $2p2h$ interactions. 

The SuSAv2 model~\cite{SuSAv2QE} is based on a superscaling model that accurately describes QE scattering on a variety of targets ($A \ge 12$) for a wide range of electron energies. The version used in GENIE~\cite{Dolan:2019bxf} describes struck nucleon momenta with a mean field model (similar to LFG) at low momentum transfer and with a relativistic Fermi Gas at high momentum transfer.  It was extended to describe $2p2h$ scattering using a theoretical model.  This model describes $(e,e')$ scattering data at QE kinematics~\cite{Megias:2018ujz} very well.  We used this model for QE and $2p2h$ processes because its agreement with data is somewhat better than what was obtained with the G18 model~\cite{e4nu-egenie:2020tbf}.  It was compared to $(e,e'p)$ data in Ref. \cite{e4nu-nature}.  For use in simulations of experiment, approximations were made describing the outgoing nucleon kinematics~\cite{Bodek:1980ar} using an LFG nuclear model. We then characterize the SuSAv2 nuclear model as LFG for the studies in this work. 

GENIE describes pion production using the Berger-Sehgal model~\cite{Berger-Sehgal} with an internal non-resonant model developed by Andreopoulos, Gallagher, Kehias, and Yang~\cite{Yang:2009zx} used together with a Bodek-Yang DIS model~\cite{Bodek:2005de}. 

For final state interactions, the G18 model uses the  hA2018 model~\cite{Dytman:2011zz} and the SuSAv2 model uses the hN2018 model. Both models use the Monte Carlo technique of stepping through the nuclear medium to decide when an interaction occurs. Both models apply the same stepping process for nucleons. The hN2018 model is an intranuclear cascade (INC) model based on free hadron-nucleon cross sections~\cite{said} with nuclear corrections~\cite{Salcedo:1987md,Pandharipande:1992zz}. The hA2018 model is more empirical, using hadron-nucleus data to generate the outgoing particles.

We also included several different nuclear models.  The default LFG model used for the analysis does not include the effect of nucleon-nucleon ($NN$) short-range correlations. These short-range correlations redistribute strength from low momentum (mean field) nucleons to high momentum (correlated) nucleons~\cite{CLAS:2003eih}. This effect is included in the Correlated Fermi Gas (CFG) model. A nuclear model using the spectral function, $S(E_m,\vec{p}_m)$~\cite{Betancourt:2023uxz}, has become available in GENIE for carbon. This implementation uses the correlated basis function (CBF) Spectral Function model~\cite{Benhar:1989aw,Benhar:1994hw}. This realistic spectral function includes correlations based on the Local Density Approximation and a depleted mean field region fit to $(e,e'p)$ data. 

%\begin{center}
\begin{figure}[htb!]
    \centering
   \includegraphics[width=8cm]{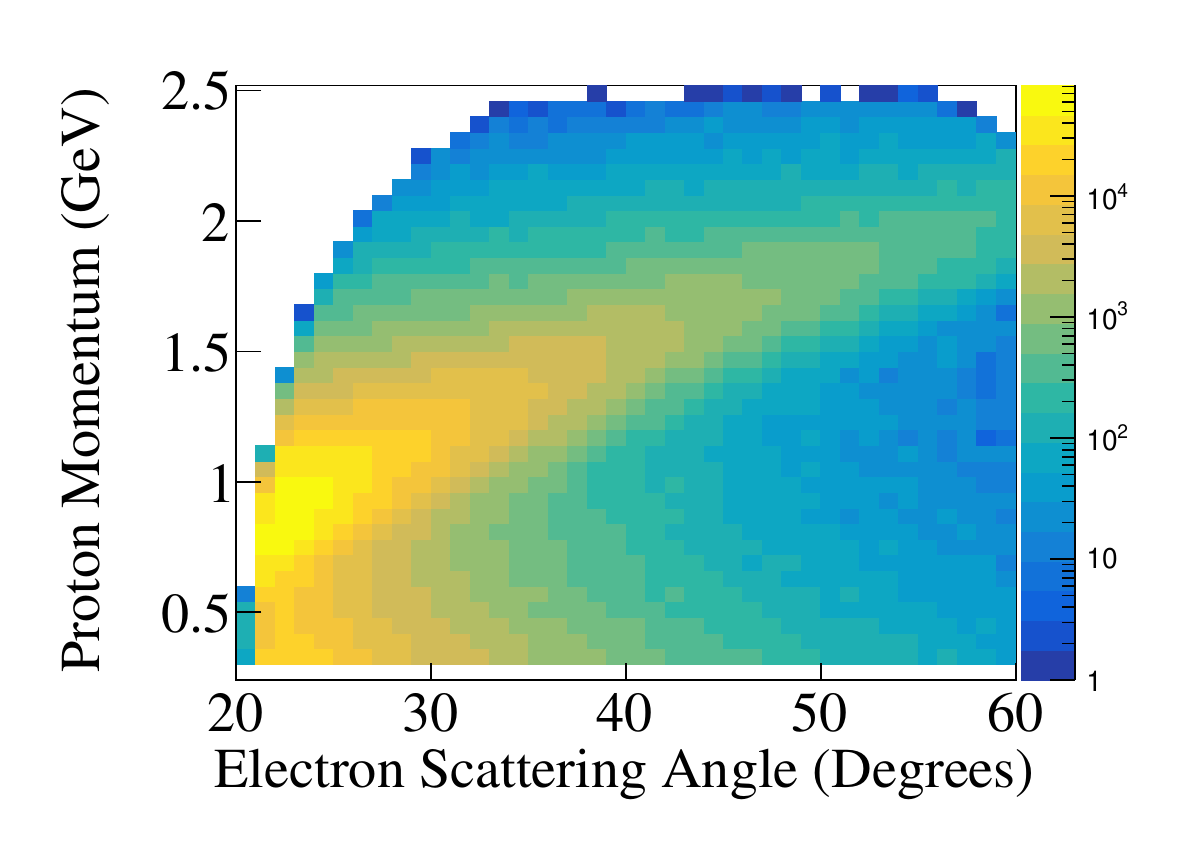}
    \caption{
     The correlation between $\theta_e$ and $p_p$ for simulated QE events for 2.261 GeV electrons incident on C.
    %Two dimensional histogram showing the correlation between electron scattering angle and proton momentum in the final state. Plot includes true QE electron-carbon scattering events at an electron beam energy of 2.261 GeV. Color indicates the number of events in arbitrary units.
    }
    \label{fig:epcorr-thmom}
    \vspace{-.5cm}
\end{figure}
%\end{center}

%The correlations in elastic scattering between electron and proton in the final state are perfect for a free proton target. In a nucleus, there is smearing due to Fermi motion and FSI. This is seen in Fig.~\ref{fig:epcorr-thmom} which shows the outgoing proton momentum ($p_p$) as a function of the outgoing electron scattering angle ($\theta_e$) for simulated QE events from electron-carbon scattering. We utilize this correlation to select bins of proton momentum by slicing the data into bins of electron scattering angle. For inclusive $(e,e')$ scattering this also allows us to separate the different reaction mechanisms in energy transfer. The final transparency results are measured as a function of $p_p$ as this is the variable against which neutrino event generators tune their FSI models. Fig.~\ref{fig:epcorr-phiphi} shows the predicted correlation between electron and proton azimuthal angle with the acceptance of the CLAS detector applied. The correlation is strong for true QE events. The smearing is less here. The recoil proton is in the opposite sector and the analysis depends on accounting for all the acceptance losses. 

%A cut is made on the electron azimuthal angle ($\phi_e$) in each of the 6 sectors so that all the correlated QE protons are accepted in the opposing sector (180 degrees away in azimuthal angle).

\section{Analysis} \label{sec:analysis}
\noindent 
We measured transparency by comparing the number of QE $(e,e'p)$ events with one detected proton, and zero pions and photons to the number of QE $(e,e')$ events that should have had a proton.  We selected the outgoing proton momentum bins by selecting bins of electron scattering angle $\theta_e$. The momentum transfer, and hence the average proton momentum, increases with electron scattering angle.  This correlation between $\theta_e$ and $p_p$ is exact for a free proton target, but is smeared by Fermi motion in a nucleus (see Fig. ~\ref{fig:epcorr-thmom}).  

We applied the same tracking, particle identification, fiducial cuts, acceptance and background subtraction as in Ref.~\cite{e4nu-nature}.  Fiducial cuts were applied to select regions of uniform and large acceptance.  Background subtraction was performed to remove events with undetected extra particles from the $(e,e'p)$ sample. This normalization canceled in the transparency ratio.

%This measurement uses data from the e2a run period of the CLAS spectrometer. Data are available for 2.261 and 4.461 GeV beam energies and helium, carbon, and iron targets. All targets have natural isotopic abundance. Therefore, the helium sample is almost totally $^4$He. The carbon comes predominantly from $^{12}$C (98.9\%) and the iron data are predominantly from $^{56}$Fe (91.8\%). The first e4$\nu$ measurement~\cite{e4nu-egenie:2020tbf} was based on these same data and many of the methods are reused here. The measurements share fiducial cuts, tracking, acceptance corrections, and background subtraction methods. The biggest difference is that the kinematic regions are broad in  Ref.~\cite{e4nu-egenie:2020tbf} and narrow here. 

\subsection{Event Selection}
\label{sec:EventSelection}

\subsubsection{$(e,e')$ Selection}
\begin{figure}[htb!]
    \centering
   \includegraphics[width=0.48\textwidth]{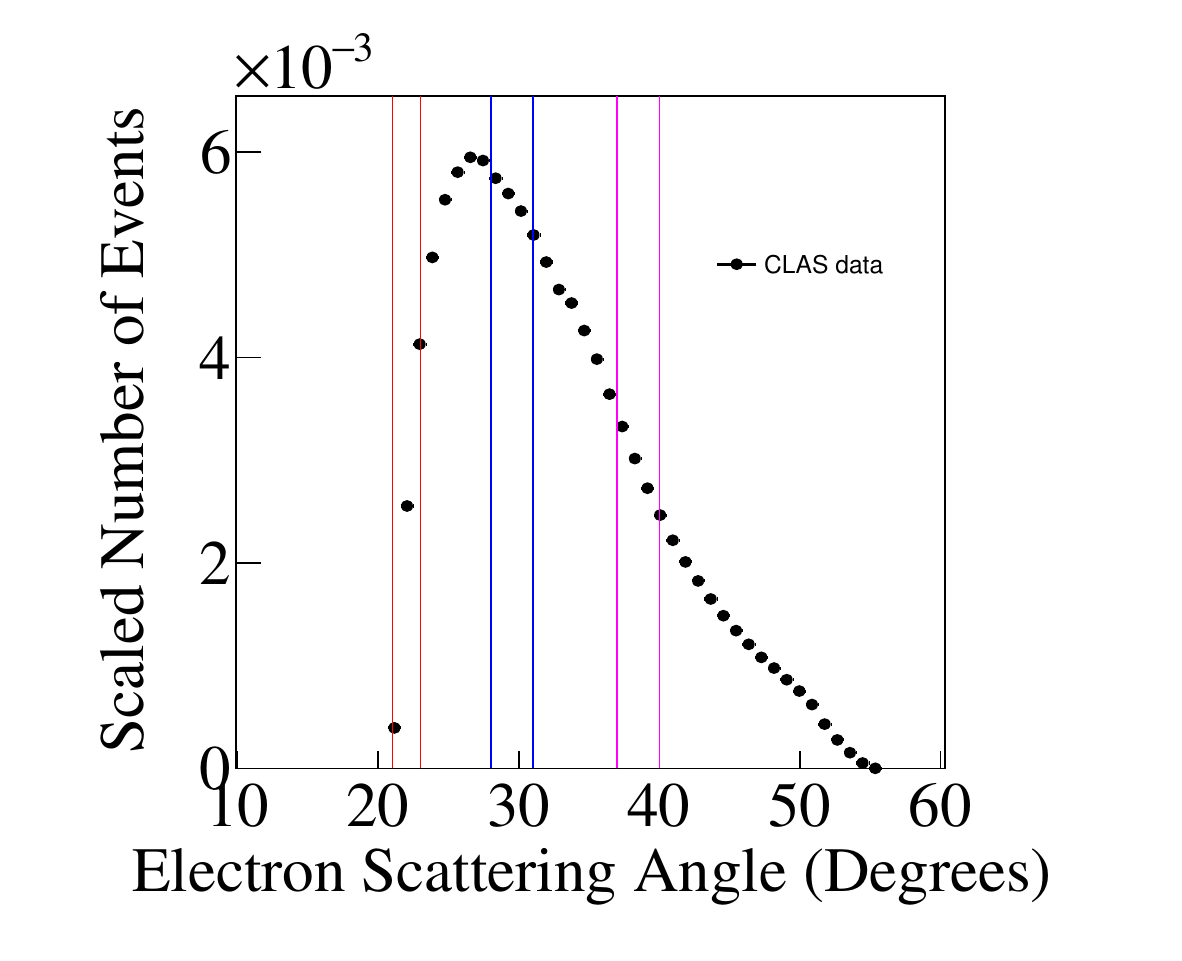}
   \includegraphics[width=0.48\textwidth]{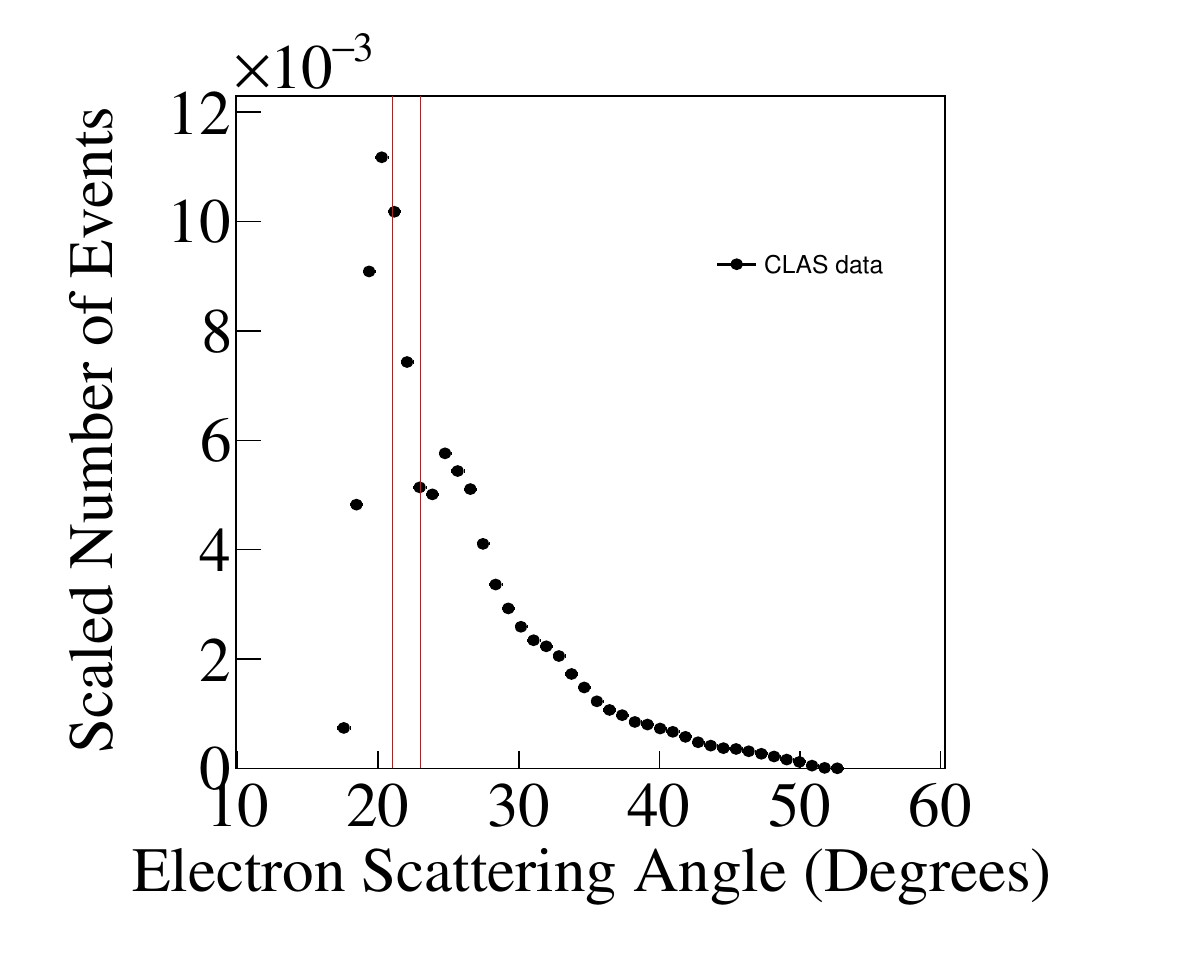}
    \caption{Top: Number of events (luminosity normalized) as a function of reconstructed electron scattering angle for 2.261 GeV carbon without cuts. Red vertical lines denote range 1 ($21^\circ<\theta_{e}<23^\circ$), blue vertical lines denote range 2 ($28^\circ<\theta_{e}<31^\circ$) degrees, and purple vertical lines denote range 3 ($37^\circ<\theta_{e}<40^\circ$). Bottom: Number of events (luminosity normalized) as a function of reconstructed electron scattering angle for 4.461 GeV Fe without cuts. Red vertical lines denote range 4 ($21^\circ<\theta_{e}<23^\circ$).}
    \vspace{-.5cm}
    \label{fig:phi_vs_thetaElectron_2}
\end{figure}

We selected bins of proton momentum by selecting three ranges of electron angles at 2.261 GeV: range 1: $21^\circ \le \theta_e\le 23^\circ$, range 2: $28^\circ \le \theta_e\le 31^\circ$, and range 3: $37^\circ \le \theta_e\le 40^\circ$ and one range at 4.461 GeV: range 4: $21^\circ \le \theta_e\le 23^\circ$ (see Fig.~\ref{fig:phi_vs_thetaElectron_2}). The choice of these bins ensures that the average proton momentum in each electron scattering angle bin is well separated from other bins. All cuts and corrections for the (e,e') events were applied to both numerator and denominator of the transparency ratio. Note that in the following plots, the data and simulation are ''luminosity normalized'', i.e. normalized to the same flux and number of scattering centers. 

%The primary signal definition comes from the choice of electron scattering angles. This provides both a primary selection of events and the range of proton momentum for the transparency measurement as seen in Fig.~\ref{fig:epcorr-thmom}. A broad view can be seen in Fig.~\ref{fig:phi_vs_thetaElectron_2} where the $\theta_e$ regions are defined for 2.261 and 4.461 GeV data. For $2.261$ GeV data we use range 1 (21 deg$^\circ< \theta_e < 23^\circ$), range 2 ($28^\circ< \theta_e < 31^\circ$) and range 3 (37 deg $< \theta_e <$ 40 deg) to label kinematic regions of interest. For $4.461$ GeV we use only a single range, (21 deg $< \theta_e <$ 23 deg), which is labeled range 4. The choice of these bins ensures that the average proton momentum in each bin of electron scattering angle is well separated from other bins. Note that in the following plots, the data and simulation are ''luminosity normalized'' to the same flux and number of scattering centers. 
%\begin{center}
%\end{center}
%\begin{center}
\begin{figure}[htb!]
    \centering
    \includegraphics[width=0.48\textwidth]{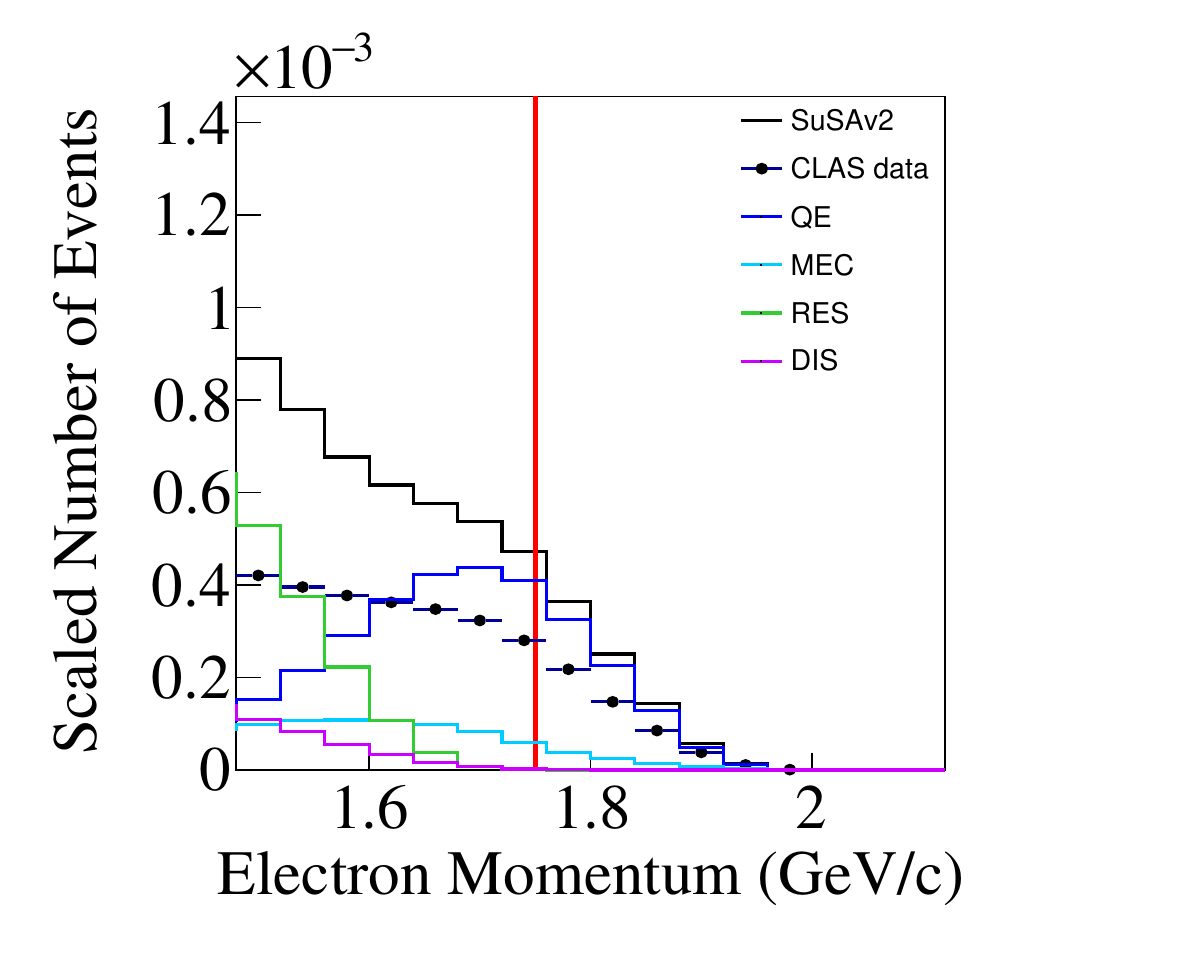}
    \includegraphics[width=0.48\textwidth]{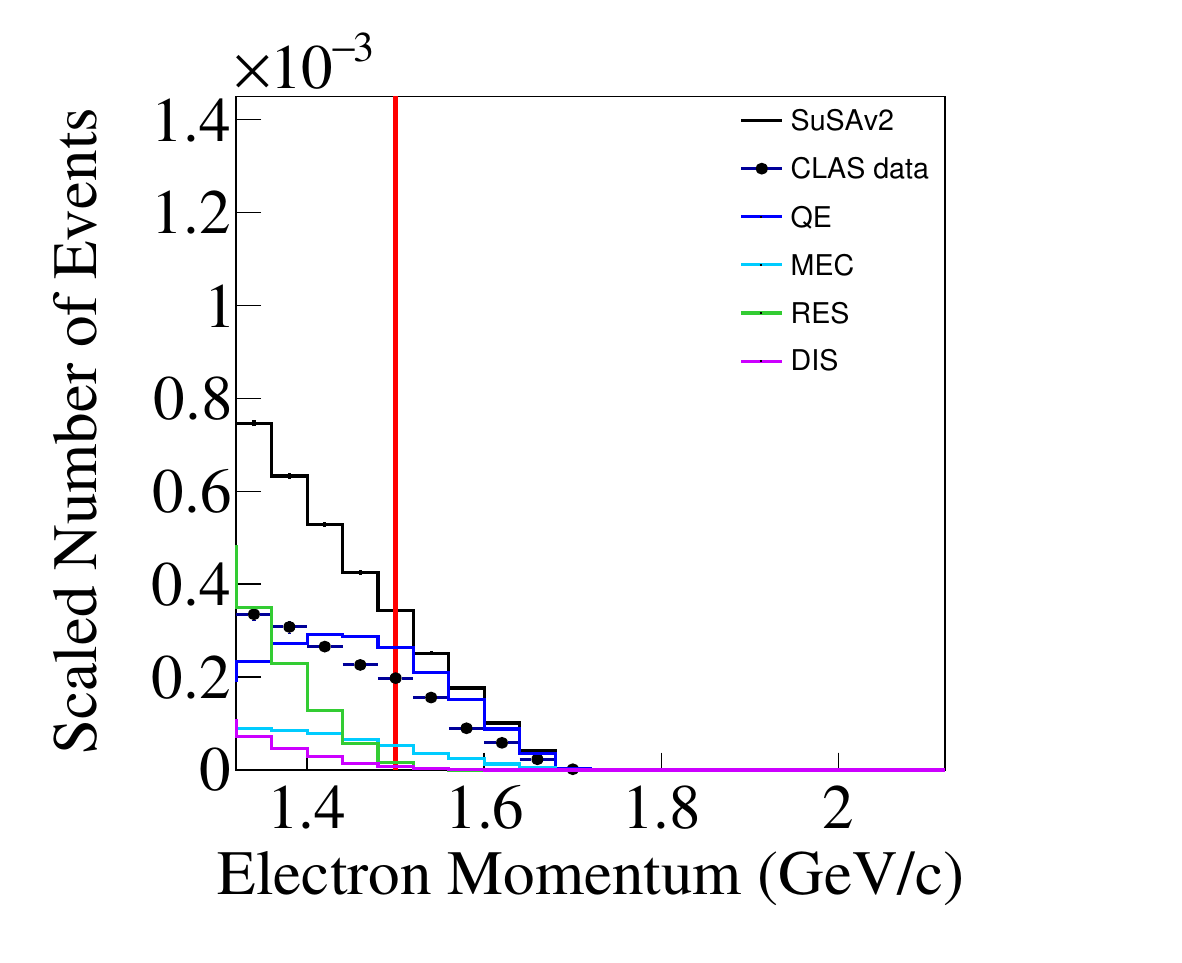}
    \caption{Top: Number of events (luminosity normalized) as a function of scattered electron momentum for beam energy 2.261 GeV for the carbon target at $28^\circ<\theta_{e}<31^\circ$ and $\Delta\phi_{e}=12^\circ$ for data (black points), GENIE (black histogram), and specific GENIE  reaction channels, QE (blue), MEC/$2p2h$ (light blue), RES (green) and DIS (purple). The vertical line denotes the electron momentum cut of 1.75 GeV/c. Bottom: Number of events (luminosity normalized) as a function of reconstructed electron momentum for 2.261 GeV Fe broken down by channel with $37^\circ<\theta_{e}<40^\circ$ and $-12^\circ \le \phi_{e} \le 12^\circ$. The vertical line denotes the electron momentum cut of 1.5 GeV/c.}
     \label{fig:elec_mom_2}
\end{figure}
%\end{center}

To maximize the fraction of true QE events in the $(e,e')$ data, we cut on the minimum electron momentum ($p_e$) for each target, energy and angular range (see Fig.~\ref{fig:elec_mom_2} and Tables I-III).  The cuts were chosen to remove most of the pion production and $2p2h$ events as predicted by GENIE. Only two $2p2h$ electron-scattering GENIE models (SuSAv2, Empirical $2p2h$) are currently available. The SuSAv2 model was chosen as the main simulation because the alternative G18 model was shown to be less accurate for both electron scattering~\cite{e4nu-nature,e4nu-egenie:2020tbf} and neutrino scattering~\cite{PhysRevLett.116.071802}. The normalization difference between data and simulation did not affect the choice of the cut.  The transparency ratio was not very sensitive to the exact value of this cut, since it was applied to both numerator and denominator and hence largely canceled.  We corrected for the remaining non-QE contamination using GENIE SuSAv2 (see Sect.~\ref{sec:CorrectionFactors}) and varied the cut to determine the associated systematic uncertainty (see Sect.~\ref{sec:systematics}). 

\begin{figure}[htb!]
   \includegraphics[width=0.47\textwidth]{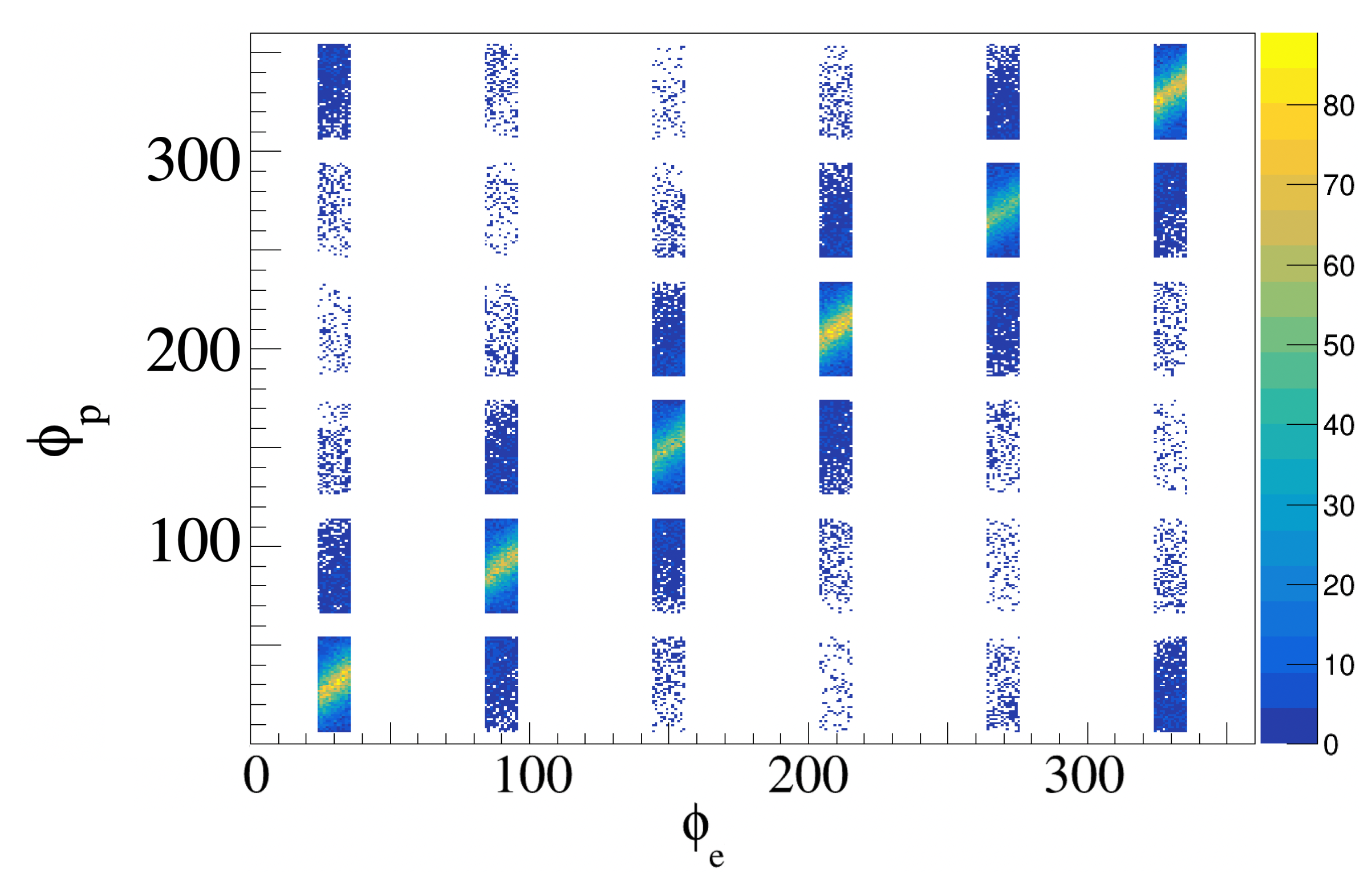}
    \caption{
    The correlation between $\phi_e$ and $\phi_p$ for simulated QE events for 2.261 GeV electrons incident on C.  The proton angle is plotted as $180^\circ -\phi_p$ so that we expect $\phi_e \approx \phi_p$.  Acceptance cuts have been applied.
    }
    \label{fig:epcorr-phiphi}
    \vspace{-.5cm}
\end{figure}
\begin{figure}[htb!]
    \centering
    \includegraphics[width=0.44\textwidth]{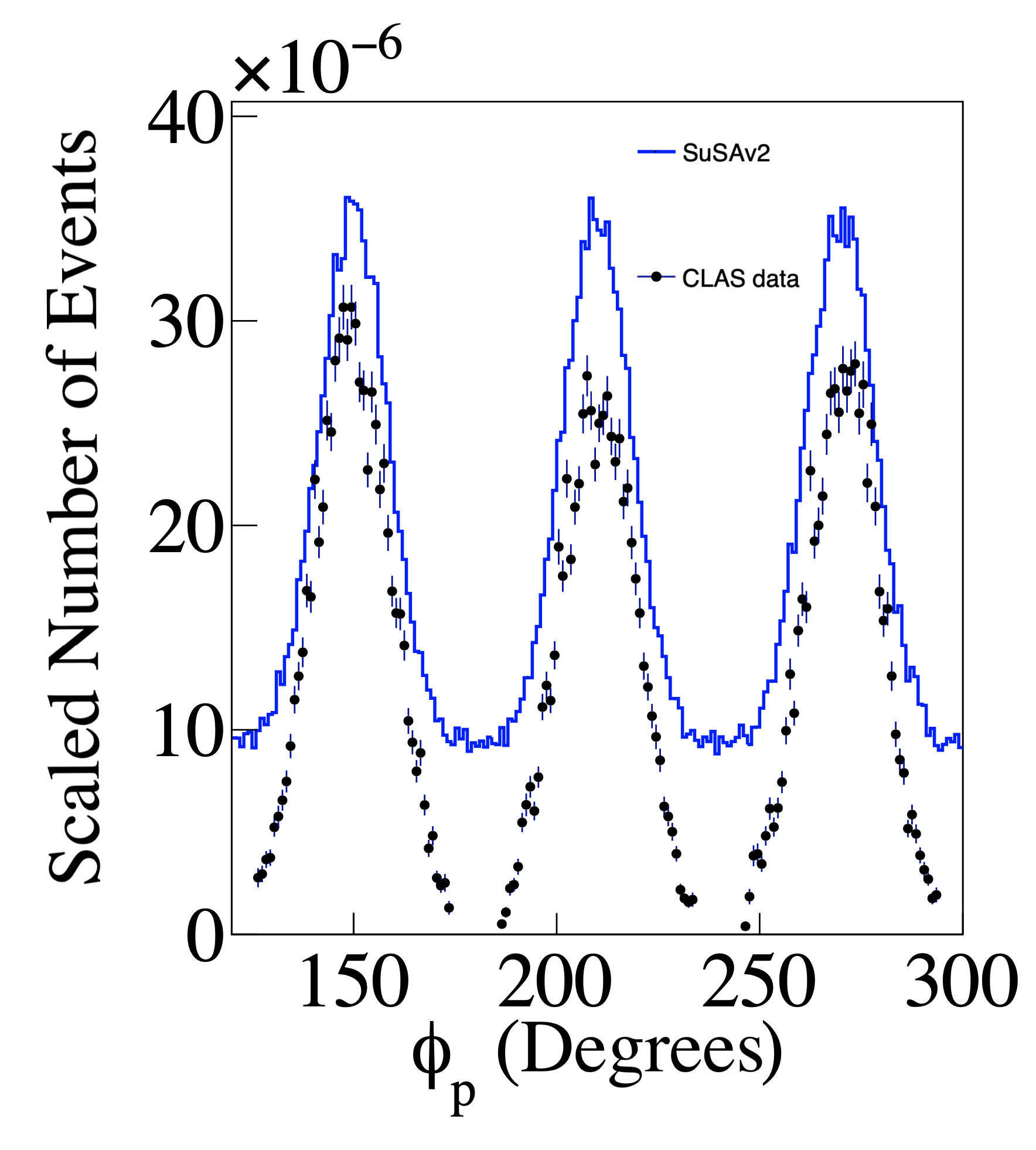}
    \caption{Number of events (luminosity normalized) as a function of proton $\phi$ for data (black points) and for generated SuSAv2 QE events (blue histogram) for 2.261 GeV C with $21^\circ<\theta_{e^-}<23^\circ$ and $-12^\circ\le\phi_{e}\le 12^\circ$. No acceptance or fiducial cuts are applied to the MC. The shape of the generated SuSAv2 spectrum is due to the tight electron angular cuts and the QE electron-proton angular correlation (see Fig.~\ref{fig:epcorr-phiphi}).  Only the three sectors used in the analysis are shown.}  
    \label{fig:phigap}
\end{figure}

\vspace{0.1in}
There is a strong correlation between the electron and proton azimuthal angles, $\phi_e$ and $\phi_p$, which is smeared by the nuclear Fermi motion (see Fig.~\ref{fig:epcorr-phiphi}).
$\phi_e$ was restricted through the $\phi_e^{max}$ parameter (the maximum allowed value) to the range $-\phi_e^{max} \le\phi_e\le \phi_e^{max}$.  This is applied in each electron sector so that the corresponding knocked-out protons would be detected in the fiducial region of the opposite CLAS sector. To check that this cut is sufficient to detect all true QE protons in the $(e,e'p)$ sample, we compared the simulated $\phi_p$ distribution without the $\phi_e$ cut with the data.  Neither data nor simulation had acceptance cuts. 
Fig.~\ref{fig:phigap} shows that these quantities have very similar shape near the center of each sector where true QE events are expected. In the regions between sectors, there are no data events because of acceptance and the simulation has only events from protons which underwent FSI. 

%\begin{center}
%\end{center}  

\subsubsection{$(e,e'p)$ Selection}

The numerator of the transparency ratio is composed of true QE events with an electron and a proton, with the proton passing all tests as having had no FSI. 
It contains both an electron that satisfies the cuts discussed in the previous section and a new set of cuts.

\begin{figure}[htb!]
    \centering
    \includegraphics[width=0.48\textwidth]{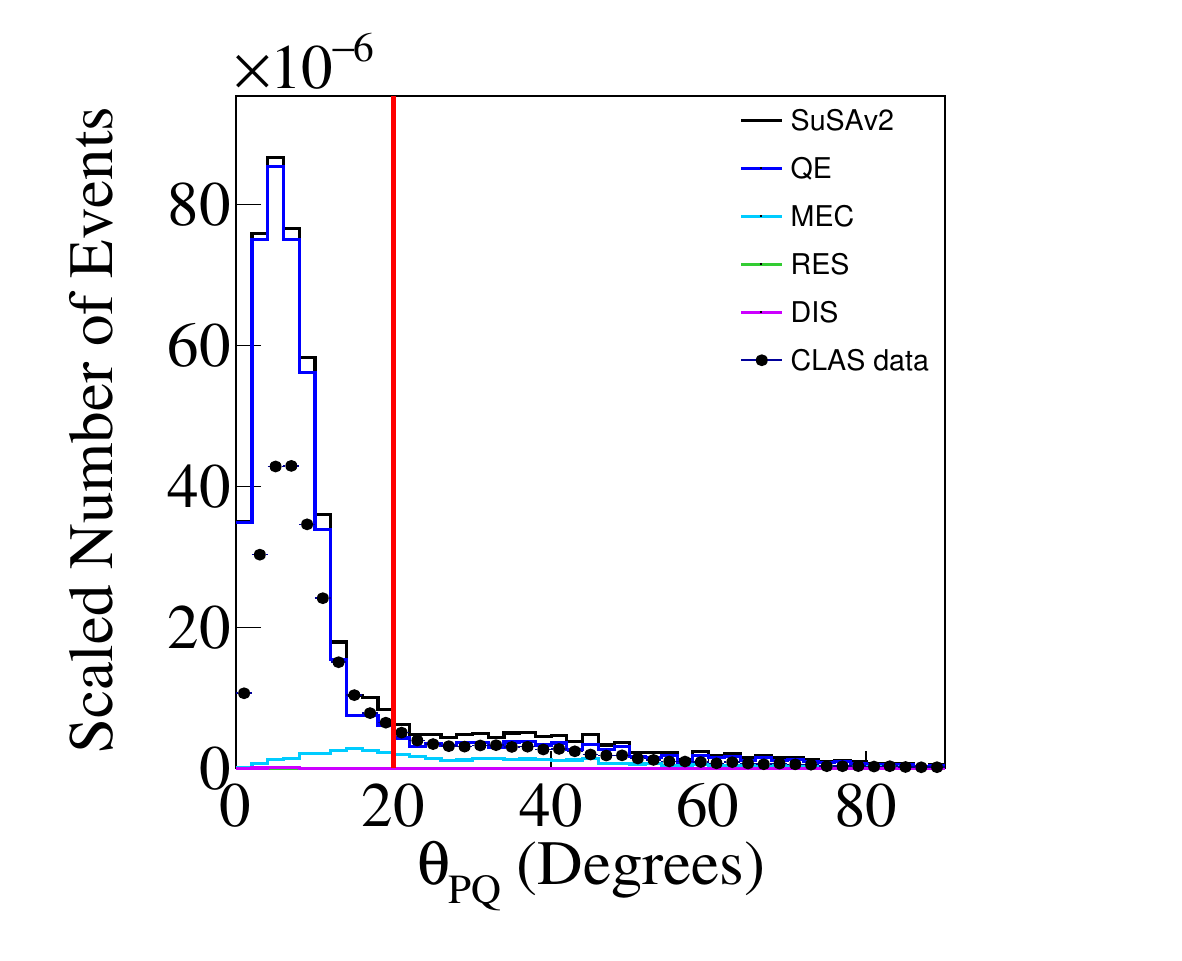}
    \includegraphics[width=0.48\textwidth]{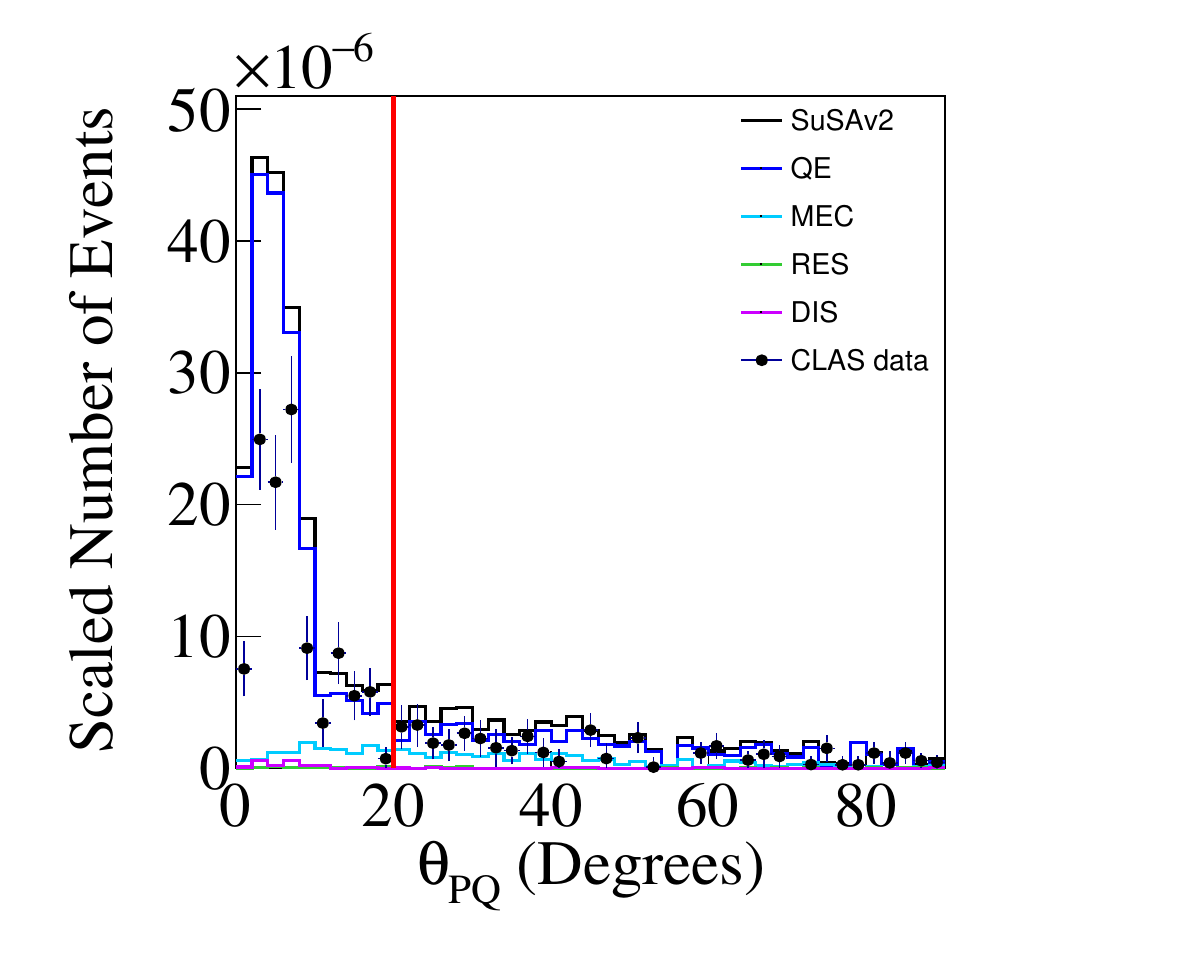}
    \caption{Top: Number of events (luminosity normalized) as a function of reconstructed  $\theta_{PQ}$ for 2.261 GeV C for $28^\circ<\theta_{e}<31^\circ$  for data (black points), GENIE (black histogram), and specific GENIE  reaction channels: QE (blue), MEC/$2p2h$ (light blue), RES (green) and DIS (purple). The vertical lines denote the $\theta_{PQ}\le 20^\circ$ cut. Bottom: The same for 2.261 GeV Fe for  $37^\circ\le\theta_{e}\le 40^\circ$.}
    \vspace{-.5cm}
     \label{fig:thetaPQ_C12}
\end{figure}

In order to minimize the non-QE and FSI contributions, we require $\theta_{PQ}\le 20^\circ$, where $\theta_{PQ}$ corresponds to the angle between the virtual photon momentum ($\vec{q}$) and the final state proton momentum  ($\vec{p}_p$). This quantity is often used to isolate QE events in electron scattering experiments~\cite{CLAS:2018xsb}. Figure~\ref{fig:thetaPQ_C12} shows peaks at small $\theta_{PQ}$ corresponding to unrescattered QE protons and a tail extending to larger $\theta_{PQ}$ for rescattered and non-QE protons.  The peak is broadened by nuclear Fermi motion. The qualitative features are the same for both data and simulation.
%\begin{center}
%\end{center}

\begin{figure}[htb!]
    \centering
        \includegraphics[width=0.53\textwidth]{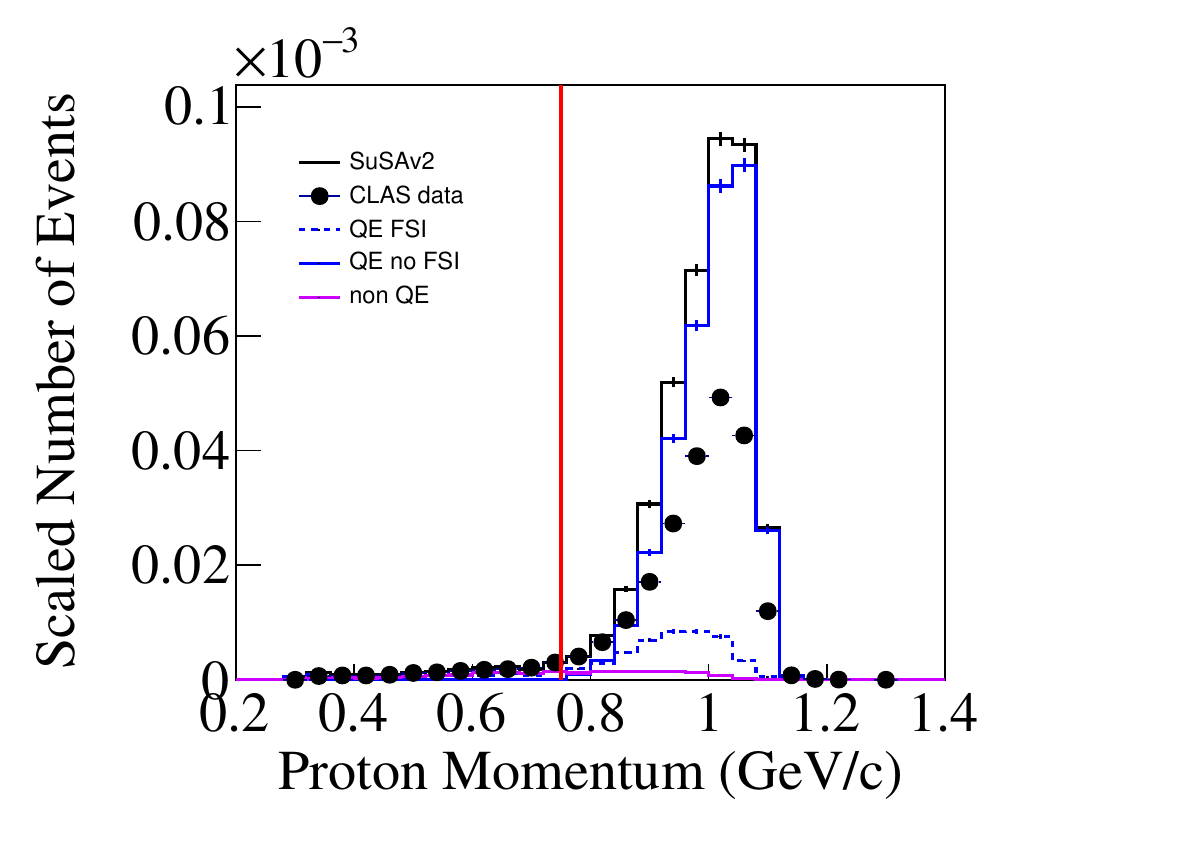}
        % 8_9_ProtMomcut_C12_2.261000_2.pdf}
        \includegraphics[width=0.53\textwidth]{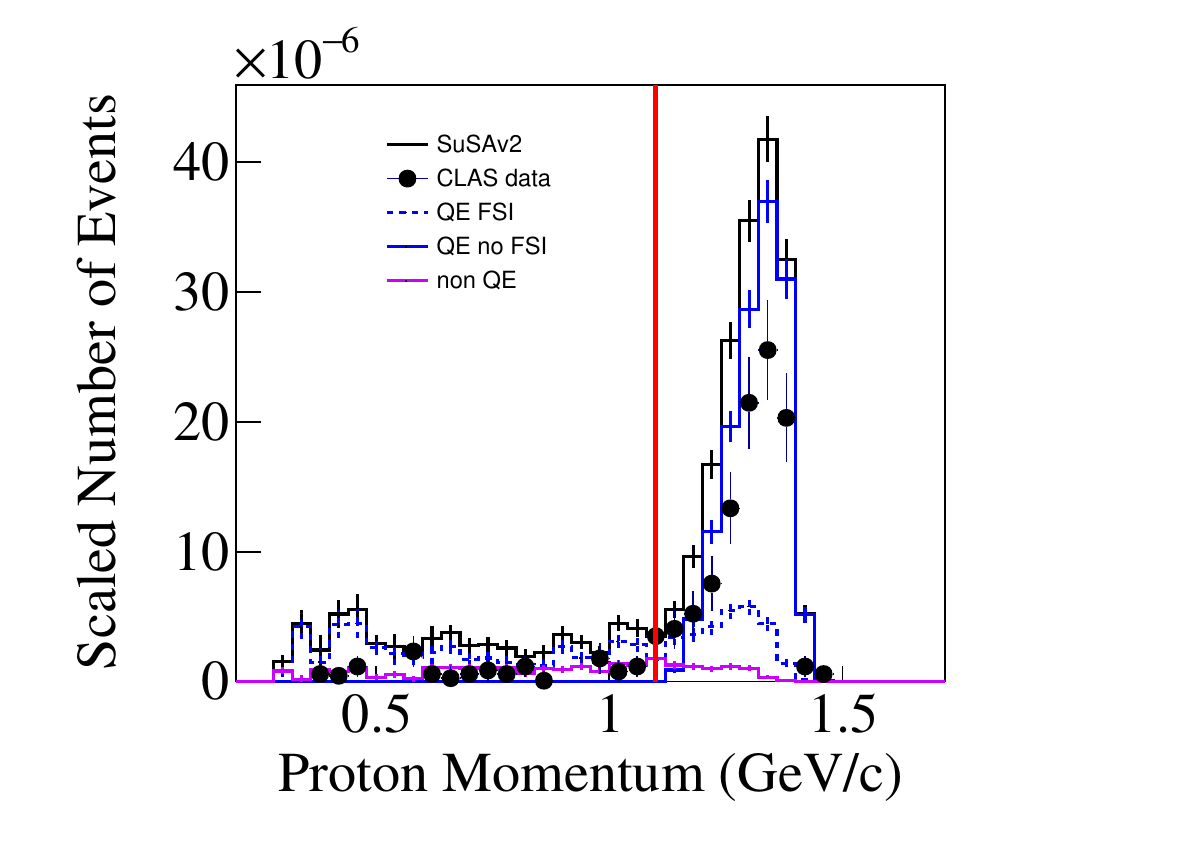}
        \caption{Top: Number of events (luminosity normalized) as a function of reconstructed proton momentum for 2.261 GeV carbon with $28^\circ\leq\theta_e\leq31^\circ$.  The black points show the data, the black histogram shows the total GENIE SuSAv2 result, the solid blue histogram shows the QE contribution for protons that did not rescatter and the dashed blue histogram shows the QE contribution for protons that did  rescatter.  The magenta line shows the non-QE contribution. 
        Bottom: Same for the 2.261 GeV iron target with $37^\circ\leq\theta_e\leq40^\circ$.}
        \label{fig:prot-mom-c-r1}
\end{figure}

Figure~\ref{fig:prot-mom-c-r1} shows the $(e,e'p)$ proton momentum distributions after the $p_e$ and $\theta_{PQ}$ cuts. The main feature in both data and simulation is the peak for true QE events and the tail at lower $p_p$ that is due to non-QE and FSI events. 
%as well as a small number of inelastic events.  SD: WAS THIS CONFUSING? 
We cut on proton momentum to remove these events.  There is an additional contribution to the simulated distribution within the cut coming from small momentum transfer rescattering processes. These protons are tagged as {\it interacting} by MC even though they should have been reabsorbed into the residual nucleus because of Pauli blocking. GENIE FSI hA and hN models don't have this effect~\cite{Dytman:2021ohr}. 
However, they still pass all cuts and are correctly analyzed by MC as {\it non-interacting}.  
%STEVE However, the cuts applied to the simulation correctly labels these events as those for which no FSI has occurred and thus the transparency calculation is still correct.
%\begin{center}
%\end{center}
%\vskip -0.36in

%\FloatBarrier
\subsection{Acceptance}
\label{sec:acceptance}
The CLAS detector has large acceptance in angle and momentum.  However, due to the narrowness of the electron angular cuts, dead detector channels can have a disproportionate effect on some of the defined ranges.  We calculated the detector efficiency as a function of momentum and angle using the CLAS GEANT3 simulation Ref.~\cite{e4nu-nature}.  We corrected each event using the corresponding efficiency.  We then eliminated sectors from the analysis for each $\theta_e$ range if the efficiency was less than 95\%.  Protons were detected in the opposite sector from the electron (e.g., an electron detected in sector 1 must have the associated proton in sector 4) and were subject to a similar selection. The resulting sectors are shown in Tab. \ref{tab:cuts-he-2.2}-\ref{tab:cuts-fe-2.2}.

\begin{table}[htb!]
\centering
\begin{tabular}{|c||c|c|c|c|}
 \hline
 \multirow{3}{*}{Cuts} & \multicolumn{3}{c|}{$2.261$ GeV} & $4.461$ GeV\\
\cline{2-5} 
 & \multicolumn{4}{c|}{$\theta_e$ Ranges (degrees)}\\
\cline{2-5}
 & 21-23 & 28-31 & 37-40 & 21-23\\
 \hline\hline
 Electron Sectors* & 1,4,6 & 1,3,4,6 & 3,4,6 & 1,2,6\\\hline
 $\phi_e^{max}$ (degrees)* & 12 & 12 & 12 & 12 \\\hline
 Min $p_e$ (GeV/c)* & 1.85 & 1.75 & 1.5 & 3.4\\\hline
 Proton Sectors & 4,1,3 & 4,6,1,3 & 6,1,3 & 4,5,3\\\hline
 Max $\theta_{PQ}$ (degrees) & 20 & 20 & 20 & 20\\\hline
 Min $p_{p}$ (GeV/c) & 0.5 & 0.7 & 1.0 & 1.35\\\hline
 \hline
\end{tabular}

\vspace{0.1in}

\caption{Cut values for helium data at electron energies of 2.261 and 4.461 GeV. Cuts that are applied to both numerator and denominator are labeled with a *. Other cuts are only applied to the numerator.}
\label{tab:cuts-he-2.2}
\end{table}
\begin{table}[htb!]
\centering
\begin{tabular}{|c||c|c|c|c|}
 \hline
 \multirow{3}{*}{Cuts} & \multicolumn{3}{c|}{$2.261$ GeV} & $4.461$ GeV\\
\cline{2-5} 
 & \multicolumn{4}{c|}{$\theta_e$ Ranges (degrees)}\\
\cline{2-5}
 & 21-23 & 28-31 & 37-40 & 21-23\\
 \hline\hline
 Electron Sectors* & 1,2,6 & 1,2,3,4,6 & 3,4,6 & 1,5,6\\\hline
 $\phi_e^{max}$ (degrees)* & 12 & 12 & 12 & 12 \\\hline
 Min $p_e$ (GeV/c)* & 1.95 & 1.75 & 1.5 & 3.4\\\hline
 Proton Sectors & 4,5,3 & 4,5,6,1,3 & 3,6,1 & 4,5,2\\\hline
 Max $\theta_{PQ}$ (degrees) & 20 & 20 & 20 & 20\\\hline
 Min $p_{p}$ (GeV/c) & 0.5 & 0.75 & 1.1 & 1.4\\\hline
 \hline
\end{tabular}
\caption{Cut values for carbon data at electron energies of 2.261 and 4.461 GeV. Cuts that are applied to both numerator and denominator are labeled with a *. Other cuts are only applied to the numerator.}
\label{tab:cuts-c-2.2}
\end{table}
\begin{table}[htb!]
\centering
\begin{tabular}{|c||c|c|c|c|}
 \hline
 \multirow{3}{*}{Cuts} & \multicolumn{3}{c|}{$2.261$ GeV} & $4.461$ GeV\\
\cline{2-5} 
 & \multicolumn{4}{c|}{$\theta_e$ Ranges (degrees)}\\
\cline{2-5}
 & 21-23 & 28-31 & 37-40 & 21-23\\
 \hline\hline
 Electron Sectors* & 1,3,5,6 & 1,3,4,6 & 1,3,4,6 & 1,2,6\\\hline
 $\phi_e^{max}$ (degrees)* & 12 & 12 & 12 & 12 \\\hline
 Min $p_e$ (GeV/c)* & 1.95 & 1.75 & 1.5 & 3.35\\\hline
 Proton Sectors & 4,6,2,3 & 4,6,1,3 & 4,6,1,3 & 4,5,3\\\hline
 Max $\theta_{PQ}$ (degrees) & 20 & 20 & 20 & 20\\\hline
 Min $p_{p}$ (GeV/c) & 0.5 & 0.8 & 1.1 & 1.4\\\hline
 \hline
\end{tabular}
\caption{Cut values for iron data at electron energies of 2.261 and 4.461 GeV. Cuts that are applied to both numerator and denominator are labeled with a *. Other cuts are only applied to the numerator.}
\label{tab:cuts-fe-2.2}
\end{table}

This measurement selected events where there is only 1 proton (or the potential for 1 proton) in the final state.
The main background originated with events where more than 1 nucleon or any number of pions are in the final state.  If the extra particles were in the detector active area, they would be recorded as background.  Events with undetected extra particles (e.g.,  in the gaps between the sectors) were subtracted by a procedure developed for the previous $e4\nu$ measurement~\cite{e4nu-nature}.  Background events with multiple nucleons or any number of pions 
that were selected were rotated around the momentum transfer vector and used to estimate the number of events with undetected particles.  %The events containing these extra particles are considered to be the same independent of whether they are detected or not.  
%That is true except for a small contribution from interference response functions.  A systematic uncertainty for these effects is included and discussed in Sec.~\ref{sec:systematics}. 
Because the numerator of the transparency ratio required a very tight kinematic correlation of electron and proton, background events with more than one proton were very unlikely to fall within those cuts.

\subsection{Correction Factors}
\label{sec:CorrectionFactors}
Although the cuts in Sect.~\ref{sec:EventSelection} isolated mostly true QE events, some corrections were required. The $(e,e')$ denominator also contained true QE events where the struck nucleon was a neutron instead of a proton. We corrected for this using the well-known cross sections for $ep$ and $en$ elastic scattering, multiplying the $(e,e')$ yield by the resulting factors (see Tab.~\ref{tab:n_correction}). We used the GENIE G18 model for this purpose because the version of SuSAv2 used included only isoscalar contributions to QE scattering. An additional correction factor was needed for SuSAv2 to have the correct ratio of $ep$ to $en$ scattering.

A small amount of MEC/$2p2h$ background remained after all the cuts. We estimated this contamination using the GENIE SuSAv2 model separately for the $(e,e')$ and $(e,e'p)$ samples. These are applied as multiplicative factors to each sample (see Tab.~\ref{tab:mec_correction}). 

\begin{table*}[htb]
\centering
 \begin{tabular}{|c||c|c|c|c|}
 \hline
\multirow{2}{*}{Nucleus} & \multicolumn{3}{c|}{$2.261$ GeV} & $4.461$ GeV\\
\cline{2-5}
    & $21^\circ - 23^\circ$ & $28^\circ - 31^\circ$ & $37^\circ- 40^\circ$ & $21^\circ - 23^\circ$ \\\hline
   He  & 0.755 & 0.708 & 0.690 & 0.688  \\\hline
   C& 0.759 & 0.720 & 0.685 & 0.697 \\\hline
   Fe & 0.725 & 0.672 & 0.665 & 0.629  \\\hline
 \end{tabular}
\caption{Neutron $(e,e')$ correction factors calculated using GENIE G18 and applied to inclusive $(e,e')$ events.}
\label{tab:n_correction}
\end{table*}
\begin{table*}[htb]
\centering
 \begin{tabular}{|c||c|c|c|c|}
 \hline
\multirow{2}{*}{Nucleus} & \multicolumn{3}{c|}{$2.261$ GeV} & $4.461$ GeV\\
\cline{2-5}
    & $21^\circ - 23^\circ$ & $28^\circ - 31^\circ$ & $37^\circ- 40^\circ$ & $21^\circ - 23^\circ$ \\\hline
\multicolumn{5}{|c|}{Inclusive $(e,e')$}\\\hline
   He & 0.985 & 0.985 & 0.976 & 0.909  \\\hline
   C & 0.921 & 0.895 & 0.888 & 0.844 \\\hline
   Fe & 0.909 & 0.887 & 0.871 & 0.837  \\\hline
\multicolumn{5}{|c|}{Exclusive $(e,e'p)$}\\\hline
   He & 0.99 & 0.99 & 0.99 & 0.99  \\\hline
   C & 0.98 & 0.97 & 0.97 & 0.96 \\\hline
   Fe & 0.97 & 0.97 & 0.96 & 0.94  \\\hline
 \end{tabular}
\caption{$2p2h$ correction factors calculated using SuSAv2 and applied to inclusive $(e,e')$ and exclusive $(e,e'p)$ data.}
\label{tab:mec_correction}
\end{table*}
\begin{table*}[htb]
\centering
 \begin{tabular}{|c||c|c|c|c|}
 \hline
\multirow{2}{*}{Nucleus} & \multicolumn{3}{c|}{$2.261$ GeV} & $4.461$ GeV\\
\cline{2-5}
    & $21^\circ - 23^\circ$ & $28^\circ - 31^\circ$ & $37^\circ- 40^\circ$ & $21^\circ - 23^\circ$ \\\hline
\multicolumn{5}{|c|}{Transparency}\\\hline
   He & 1.045 & 1.023 & 1.023 & 1.023  \\\hline
   C & 1.06 & 1.03 & 1.03 & 1.03 \\\hline
   Fe & 1.07 & 1.033 & 1.033 & 1.033  \\\hline
 \end{tabular}
\caption{SRC correction factors applied to the final transparencies separated by nucleus, electron scattering angle range, and electron beam energy. The SRC corrections are applied as a multiplicative factor.}
\label{tab:src_correction}
\end{table*}

\begin{figure}[htb!]
    \centering
    \includegraphics[width=0.48\textwidth]{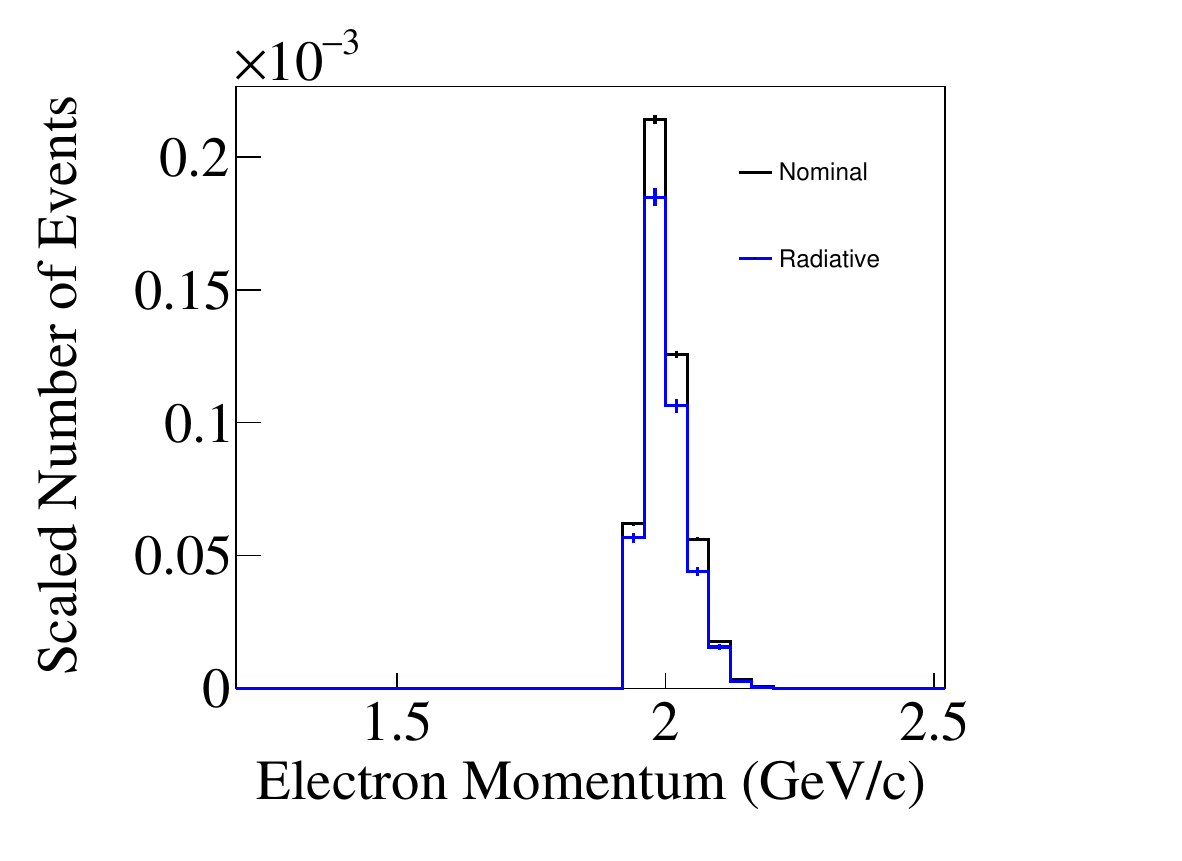}
    \caption{Number of events (luminosity normalized) vs. electron momentum for simulated 2.261 GeV $(e,e')$ events for $21^\circ<\theta_{e}<23^\circ$ detected in sector 1.  Simulations shown are the nominal SuSAv2 sample (black) without radiative corrections and the SuSAv2 radiative sample (blue).}
    \label{fig:rad_C12_incl-e}
\end{figure}

\begin{figure}[htb!]
    \centering
    \includegraphics[width=0.48\textwidth]{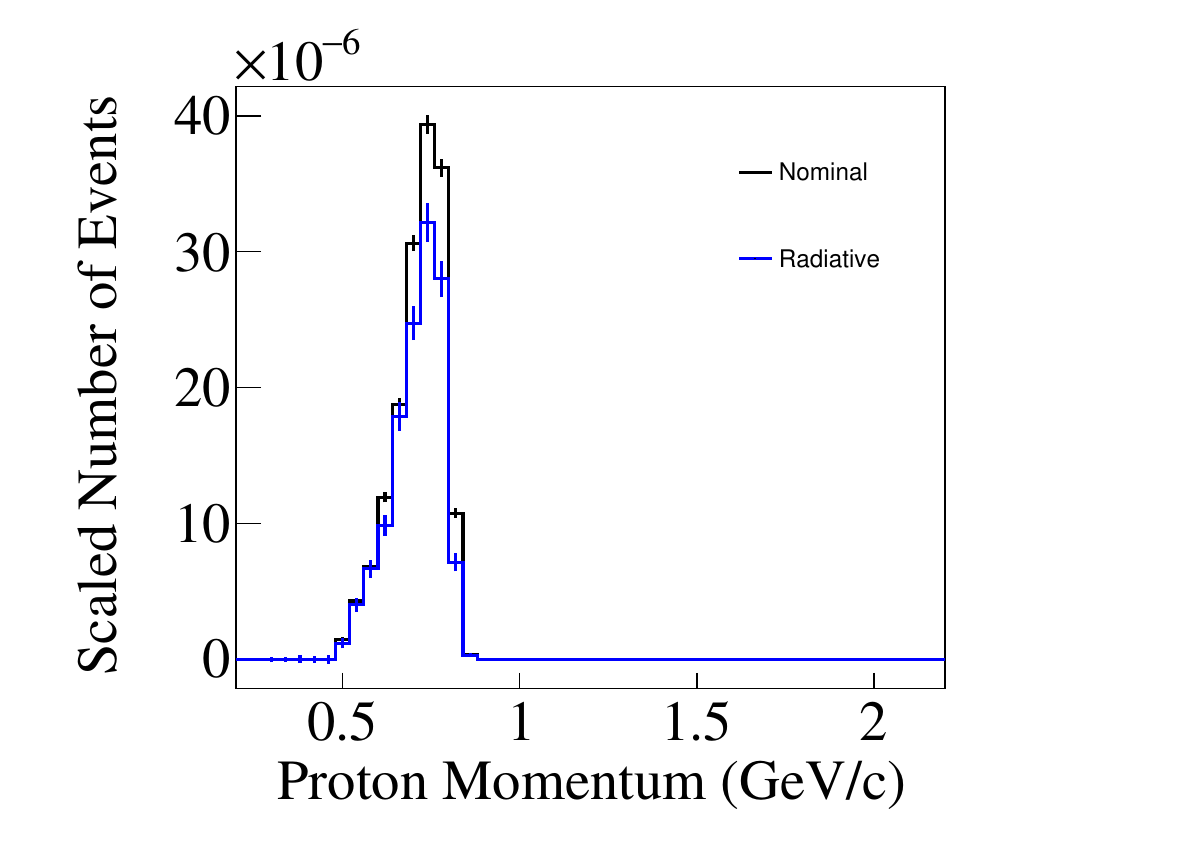}
    \caption{Number of events (luminosity normalized) vs. proton momentum for simulated 2.261 GeV $(e,e'p)$ events for $21^\circ<\theta_{e}<23^\circ$ with protons detected in sector 4. Simulations shown are the nominal SuSAv2 sample (black) without radiative corrections and the SuSAv2 radiative sample (blue)}
    \label{fig:rad_C12_incl-p}
\end{figure}

In addition, data in any electron scattering measurement must be corrected for radiative effects including vertex corrections and electron bremsstrahlung in the nuclear Coulomb field. An alternative GENIE SuSAv2 model~\cite{e4nu-egenie:2020tbf} calculated the effect of these radiative corrections for QE events using the Mo and Tsai formalism~\cite{Mo:1968cg}. This correction was applied bin-by-bin to both the $(e,e')$ and $(e,e'p)$ samples. Radiative effects largely cancel in the transparency which is a ratio of cross sections.  Figures~\ref{fig:rad_C12_incl-e} and \ref{fig:rad_C12_incl-p} show the radiatively corrected and nominal samples as a function of electron and proton momentum respectively for carbon at 2.261 GeV with electrons in sector 1 and protons in sector 4. The ratio of nominal to radiative sample as a function of electron and proton momentum on a bin-by-bin basis which was used as a correction for the data.

We also corrected the extracted transparency for the fraction of events in which an electron scatters from a nucleon belonging to a short range correlated nucleon-nucleon pair. These are pairs of nucleons with high relative momenta. Because these protons can have very large initial momentum, they might fall outside of the cuts, even for an $(e,e')$ event  in the "true QE" sample.  We calculated correction factors for these missing events using Spectral Function calculations for carbon, which compute the single nucleon knockout cross section from mean-field and correlated nucleons separately~\cite{Benhar:1994hw}, and are then extrapolated to iron and helium using experimentally measured ratios of SRC cross sections to deuterium~\cite{Hen:2012fm}. Separate correction factors were determined for the $(e,e'p)$ events. The total corrections range from 2--7\% and are listed in Tab.~\ref{tab:src_correction}. See Sect.~\ref{sec:discussion} for more details.

\subsection{Systematic Uncertainties}
\label{sec:systematics}
\begin{table*}[htb!]

\centering
\begin{tabular}{|c||c|c|c|c|c|c|c|c|} 
\hline
    Helium \% Uncertainty & Stat. & Bkgd Subt. & $2p2h$ & Accept & Cut & Sector & SRC & Total \\\hline
    $2.261$ GeV Range 1 & 0.1 & 0.34 & 0.07 & 0.11 & 0.65 & 0.1 & 2.0 & 2.14  \\
    $2.261$ GeV Range 2 & 0.26 & 0.36 & 0.07 & 0.15 & 1.36 & 1.8 & 2.0 & 3.1\\
    $2.261$ GeV Range 3 & 0.66 & 0.42 & 0.12 & 0.50 & 0.88 & 0.1 & 2.0 & 2.4  \\
    $4.461$ GeV Range 4 & 1.1 & 0.84 & 0.11 & 0.56 & 1.8 & 4.4 & 2.0 & 5.4 
    \\\hline
\end{tabular}

 \begin{tabular}{|c||c|c|c|c|c|c|c|c|} 
    Carbon \% Uncertainty & Stat. & Bkgd Subt. & $2p2h$ & Accept & Cut & Sector & SRC & Total \\\hline
    $2.261$ GeV Range 1 & 0.23 & 0.54 & 0.56 & 0.37 & 2.76 & 0.77 & 2.0 & 3.6  \\
    $2.261$ GeV Range 2 & 0.30 & 0.37 & 0.53 & 0.31 & 2.03 & 1.7 & 2.0 & 3.5\\
    $2.261$ GeV Range 3 & 1.1 & 0.49 & 0.89 & 0.56 & 2.79 & 0.1 & 2.0 & 3.8  \\
    $4.461$ GeV Range 4 & 4.0 & 1.0 & 1.1 & 0.5 & 0.83 & 7.1 & 2.0 & 8.6 
        \\\hline
 \end{tabular}
 \begin{tabular}{|c||c|c|c|c|c|c|c|c|} 
    Iron   \% uncertainty & Stat. & Bkgd Subt. & $2p2h$ & Accept & Cut & Sector & SRC & Total \\\hline
    $2.261$ GeV Range 1 & 1.4 & 0.62 & 0.74 & 0.58 & 2.83 & 4.1 & 2.0 & 5.7  \\
    $2.261$ GeV Range 2 & 2.3 & 0.64 & 1.0 & 0.41 & 3.79 & 8.03 & 2.0 & 9.4\\
    $2.261$ GeV Range 3 & 4.4 & 0.56 & 1.1 & 0.64 & 3.36 & 18.0 & 2.0 & 19.0 
    \\\hline
 \end{tabular}
 \caption{Statistical and systematic uncertainties for each nucleus in percent. This includes statistical (Stat) and systematic uncertainties due to background subtraction (Bkgd), $2p2h$ correction factors ($2p2h$), acceptance correction factors (Accept), cut variation (Cut), sector-to-sector variance (Sector), SRC Correction factors (SRC), and the total uncertainty added in quadrature.}
 \label{tab:systematics}
\end{table*}
This analysis has been designed to isolate samples of true QE scattering involving both the electron and proton. Various cuts were employed to isolate that sample and corrections were applied to account for events that remain after the cuts and are not true  QE.  Systematic uncertainties were calculated using data as much as possible and using MC calculations otherwise. All systematic and statistical uncertainties were added in quadrature to give the total uncertainty and are detailed below.

The first uncertainty is on the acceptance correction. We calculated the acceptance corrections separately using both G18 and SuSAv2.  We took the average as the acceptance correction and the difference as the uncertainty on that correction.   
This acceptance correction factor varies bin by bin. 

The most significant uncertainty comes directly from the uniformity of CLAS. The different CLAS sectors provided multiple independent transparency measurements at each scattered electron angle range and electron beam energy. Following Ref.~\cite{e4nu-nature}, we determined the systematic uncertainty for each point using the variance of the measured sector transparencies.  

There is also a systematic uncertainty due to the $2p2h$ correction. The SuSAv2 model was used and we applied a $25\%$ uncertainty to that correction factor to account for errors in the model. This uncertainty is small because the $2p2h$ correction factors are small.

Additional uncertainties from the background subtraction come from estimations of events that have undetected particles in the gaps between the sectors and for undetected photons. These systematics come from variations in the underlying pion production models and are identical to those used in Ref.~\cite{e4nu-nature}.  These uncertainties are $\le 1\%$. 

We also varied the event selection cuts to determine the systematic uncertainty due to the cut choice.  We varied the electron momentum cut by $\pm 50$ MeV. We similarly varied the proton momentum cut by $\pm 25$ MeV and the $\theta_{PQ}$ cut by $\pm 2^\circ$. These values were chosen to be somewhat larger than the CLAS resolutions. The resulting change in the transparency was divided by $\sqrt{12}$ to convert this from a flat distribution to a standard deviation.  Typical cut-variation systematic uncertainties are 1--3\%. 

We estimated the SRC correction factor uncertainty to be 2\% on the overall normalization for each transparency measurement, which is about 30--100\% of the SRC correction itself. 

The resulting statistical and systematic uncertainties are shown in Tab.~\ref{tab:systematics}. The statistical uncertainty is very small for forward angle ranges and significant for range 4. The systematic uncertainties tend to be small for helium and carbon but significant for iron even for range 1 where the statistics are excellent. The sector-to-sector variation is large in iron.

\section{Results}
\label{sec:results}

\begin{table*}[htb!]
\centering
 \begin{tabular}{|c||c|c|c|c|c|c|c|c|} 
    \hline
    \multirow{3}{*}{Nucleus} & \multicolumn{6}{c|}{$2.261$ GeV} & \multicolumn{2}{c|}{$4.461$ GeV}\\
    \cline{2-9}
    & \multicolumn{2}{c|}{$21^\circ - 23^\circ$} & \multicolumn{2}{c|}{$28^\circ - 31^\circ$} & \multicolumn{2}{c|}{$37^\circ - 40^\circ$} & \multicolumn{2}{c|}{$21^\circ - 23^\circ$} \\
    \cline{2-9}
    & $P_{p}$ (GeV/c) & T & $P_{p}$ (GeV/c) & T & $P_{p}$ (GeV/c) & T & $P_{p}$ (GeV/c) & T \\\hline
    He & 0.81 & $0.75\pm0.016$ & 1.00 & $0.71\pm0.024$ & 1.3 & $0.69\pm0.024$ & 1.65 & $0.71\pm0.05$ \\\hline
    C & 0.71 & $0.61\pm0.019$ & 0.99 & $0.63\pm0.017$ & 1.3 & $0.59\pm0.014$ & 1.64 & $0.60\pm0.05$ \\\hline
    Fe & 0.70 & $0.44\pm0.020$ & 0.99 & $0.41\pm0.034$ & 1.3 & $0.45\pm0.080$ & & \\\hline
 \end{tabular}
\caption{Data transparency and average proton momentum values for all nuclei and electron scattering angle ranges. Uncertainties include contributions from both systematic and statistical sources.}
\label{tab:transp_values}
\end{table*}

\begin{figure}[htb!]
    \centering
    \includegraphics[width=0.48\textwidth]{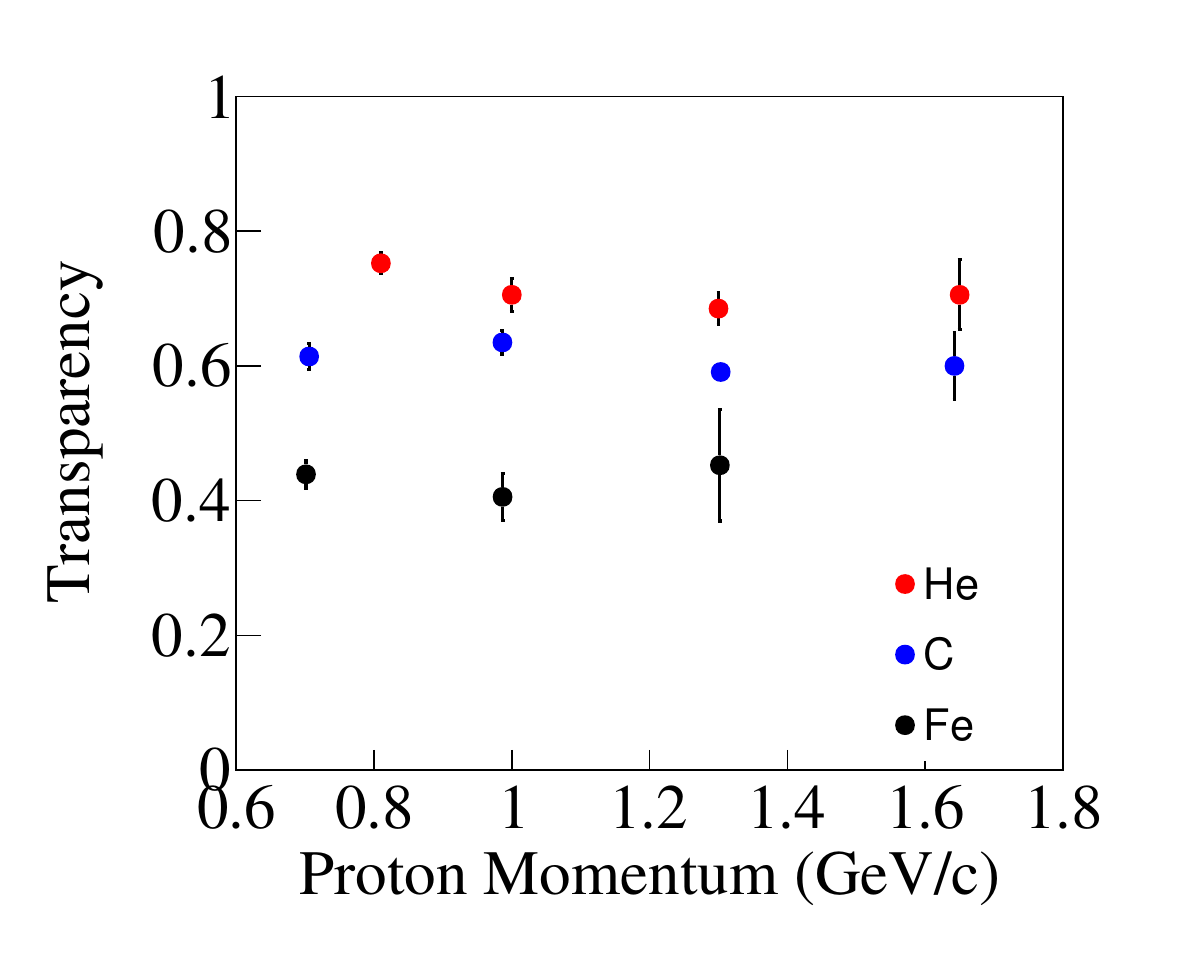}
    \caption{Transparency data as a function of proton momentum for helium (red), carbon (blue), and iron (black) from this analysis.}
        \label{fig:transp-data-all}
\end{figure}

The measured proton transparencies using the ratio of measured QE $(e,e'p)$ events to QE $(e,e')$ events for helium, carbon and iron are shown in Fig.~\ref{fig:transp-data-all} and Tab.~\ref{tab:transp_values} as a function of the average proton momentum for each of the four electron energy and angle combinations.  All cuts and corrections described in Sect.~\ref{sec:analysis} and systematic uncertainties in Sect.~\ref{sec:systematics} have been applied. Typically, systematic uncertainties are dominant except at the highest proton momenta.  However, the 4.461 GeV iron data point had very large uncertainties due to the small data sample (a few dozen events) and is therefore not shown. As expected, the transparency is largest for helium and smallest for iron, decreasing with the proton propagation distance through the nucleus.  
%{New_Plots_for_Paper/Data_Trans_newlabels.png}

\subsection{Comparison with Previous Data}
\label{sec:datacomp}
\begin{figure}[htb!]
    \centering
    \includegraphics[width=0.48\textwidth]{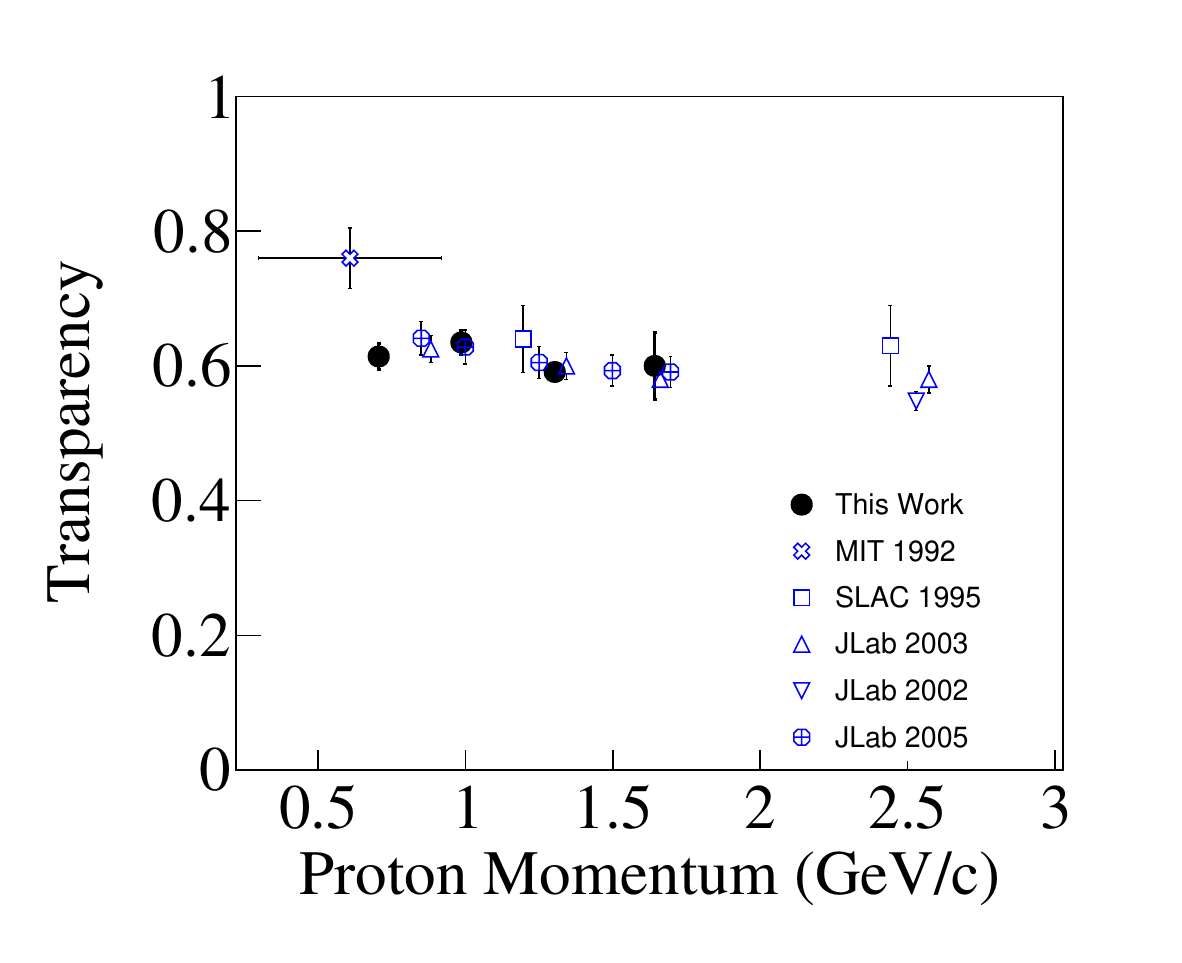}
    \caption{Transparency of carbon for data presented in this work (black filled circles) compared to world data (blue open points). Previous data are from from Jefferson Lab~\cite{Dutta:2003yt,Garrow:2001di,E97-006:2005jlg}, SLAC~\cite{ONeill:1994znv}, and MIT-Bates~\cite{Garino:1992ca}.}
      \label{fig:C_world}
\end{figure}

\begin{figure}[htb!]
    \includegraphics[width=0.48\textwidth]{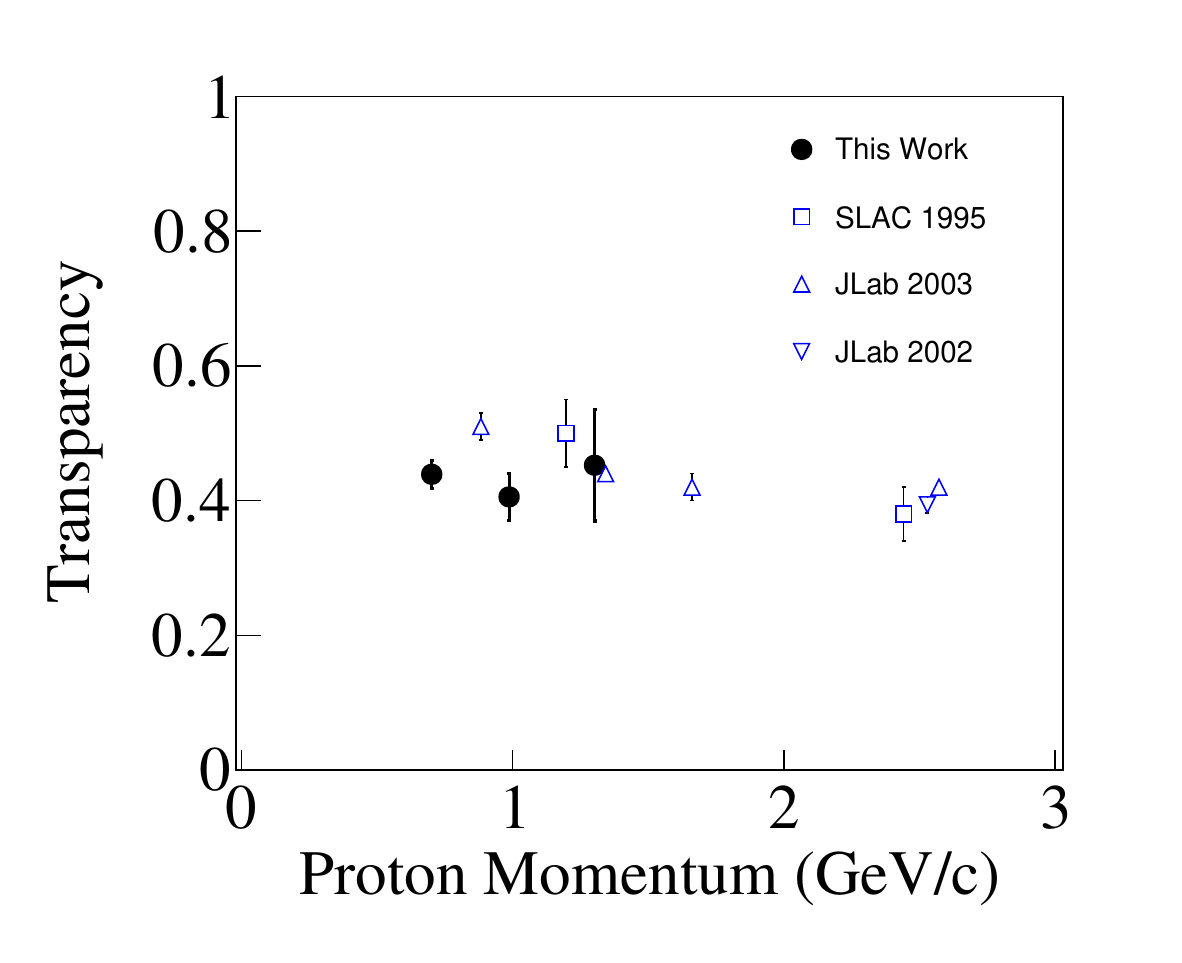}
    \caption{Transparency of iron for data presented in this work (black full circles) compared to world data (blue open points). 
 Previous data are from from Jefferson Lab~\cite{Dutta:2003yt,Garrow:2001di} and SLAC~\cite{ONeill:1994znv}.}
     \label{fig:Fe_world}
\end{figure}

We compare the data with previous transparency measurements for carbon~\cite{Garino:1992ca,ONeill:1994znv,Garrow:2001di,Dutta:2003yt,Rohe:2005vc} and iron~\cite{Dutta:2003yt,Garrow:2001di,ONeill:1994znv} in Figs.~\ref{fig:C_world} and~\ref{fig:Fe_world}, respectively. Data are provided from SLAC, Jefferson Lab, and MIT-Bates. Previous measurements typically determined  the number of QE $(e,e'p)$ events (the numerator of the transparency ratio) by cutting on $E_{miss}\le 80$ MeV and $p_{miss}\le 300$ MeV/c and determined the denominator of the ratio using a PWIA calculation of the expected number of $(e,e'p)$ events integrated over the experimental acceptance corrections, see Eq.~\ref{TAold}.  Our measurement differed in two ways.  We used cuts on $\theta_{PQ}$ and $p_p$ to select the QE $(e,e'p)$ events of the numerator and we used $(e,e')$ data (corrected for the non-QE fraction) for the denominator, see Eq.~\ref{TAnew}.  In addition, our SRC transparency correction factors are much smaller than used previously.

Our carbon transparencies are consistent with the previous results except for the point at the lowest momentum, where the previous data~\cite{Garino:1992ca} appears to trend upwards.  Our iron measurements are consistent with the average of the previous measurements at a transparency value of $\approx$0.4.  However, the previously measured transparencies increase at lower proton momentum, reaching 0.5 at $p_p\sim 1$ GeV/c, differing significantly from our measured transparencies. This increase at low momentum was also seen in other heavier nuclei~\cite{Garrow:2001di,Garino:1992ca}.

%Our measurements differ from previous ones in  our definition of transparency and our treatment of SRC.  These are discussed in Sect.~\ref{sec:discussion}.

\subsection{Comparisons with MC}
\label{sec:MCcomp}
\begin{figure}[htb!]
    \centering
        \includegraphics[width=0.48\textwidth]{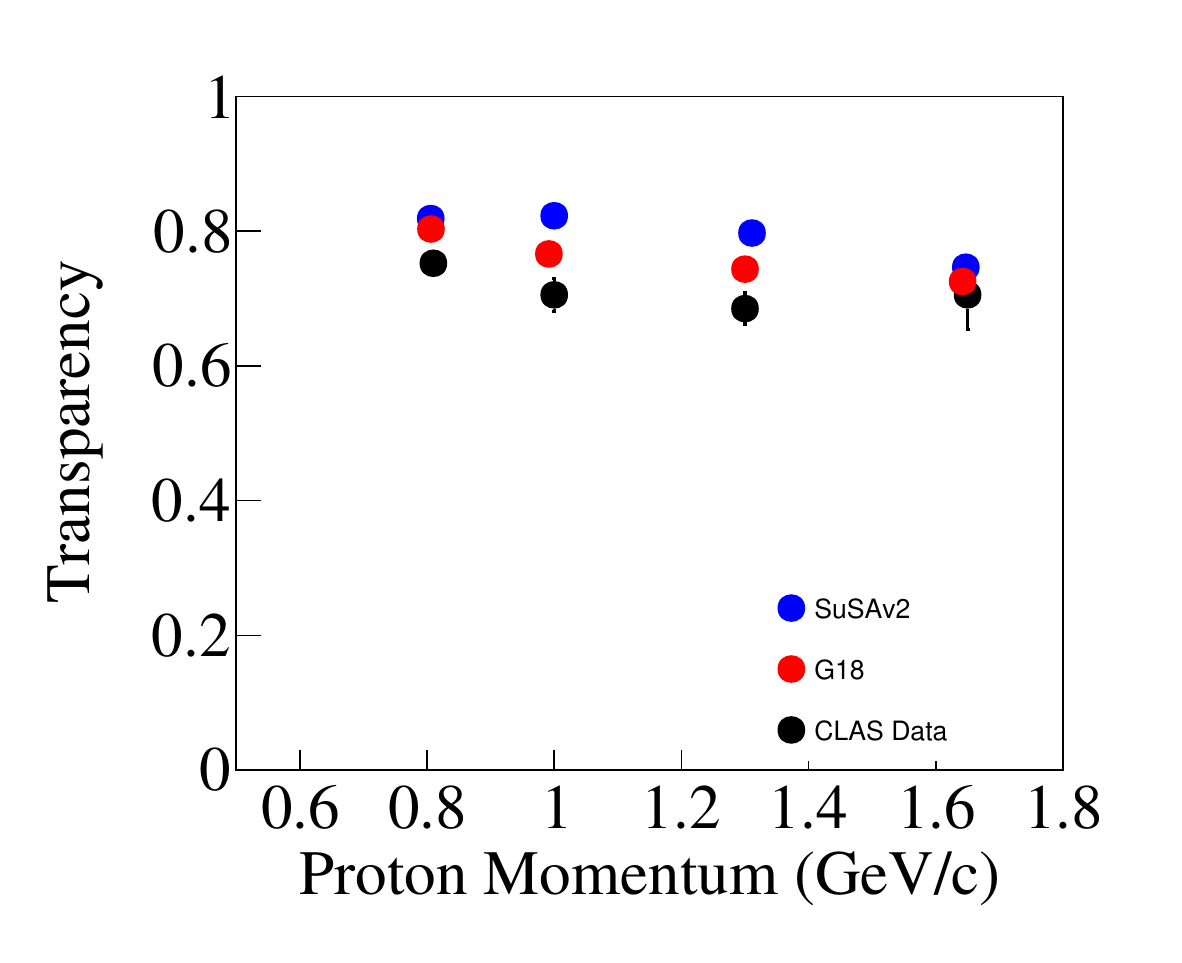}
    \caption{Proton transparency measurement for $^4$He with CLAS data (black) compared to predictions from G18+LFG+hA2018 models (red) and SuSAv2+LFG+hN2018 models (blue).}
        \label{fig:MC-data-he}
\end{figure}

\begin{figure}[htb!]
    \centering
    \includegraphics[width=0.48\textwidth] {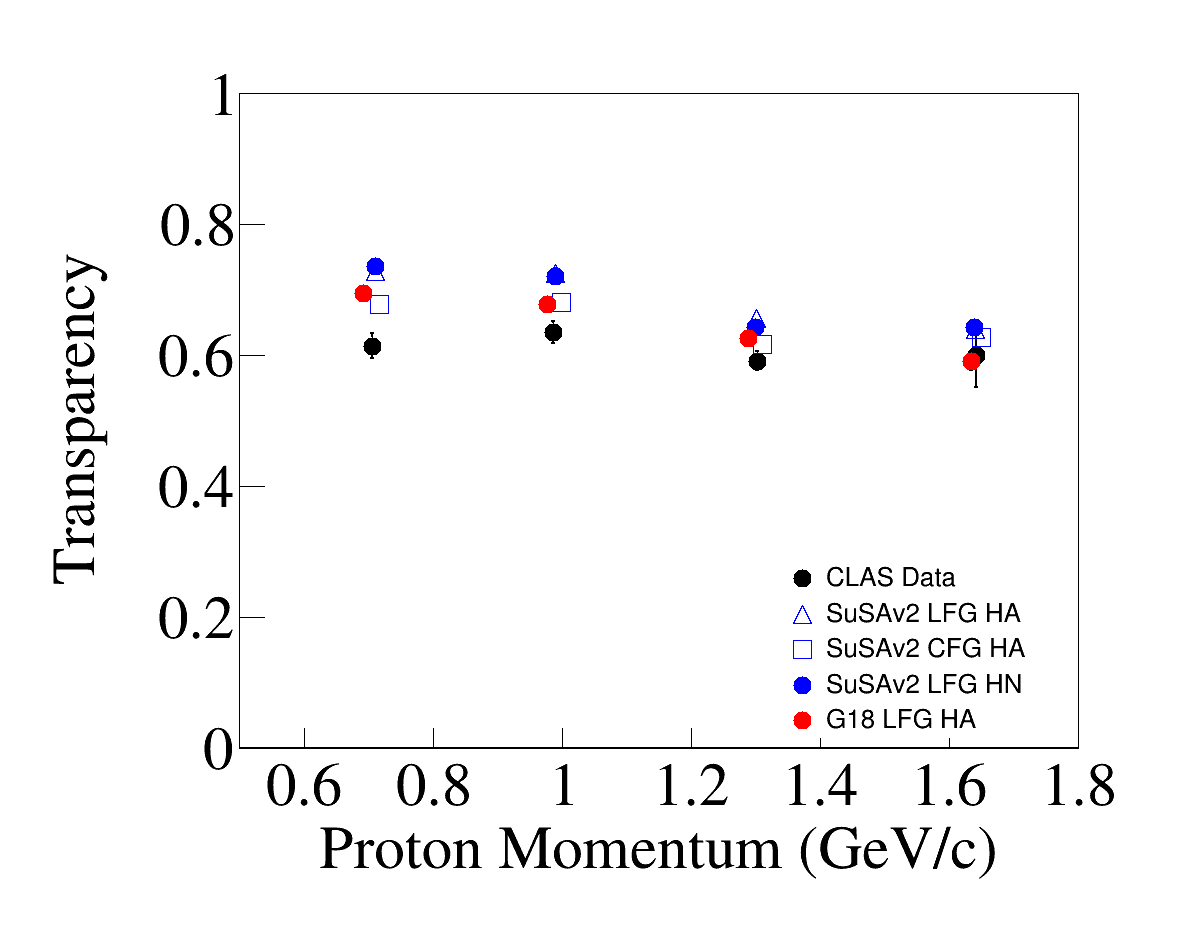}
    \caption{Proton transparency measurements for carbon with CLAS data (black) compared to various GENIE Monte Carlo predictions. These include G18+LFG+hA2018 (filled red circles) and SuSAv2+LFG+hA2018 (open blue triangles), SuSAv2+CFG+hA2018 (open blue squares), and SuSAv2+LFG+hN2018 (filled blue circles). Note SuSAv2+LFG+hN and SuSAv2+LFG+hA overlap each other so not all points are visible.}
    \label{fig:MC-data-c}
\end{figure}

\begin{figure}[htb!]
    \centering
    \includegraphics[width=0.48\textwidth]{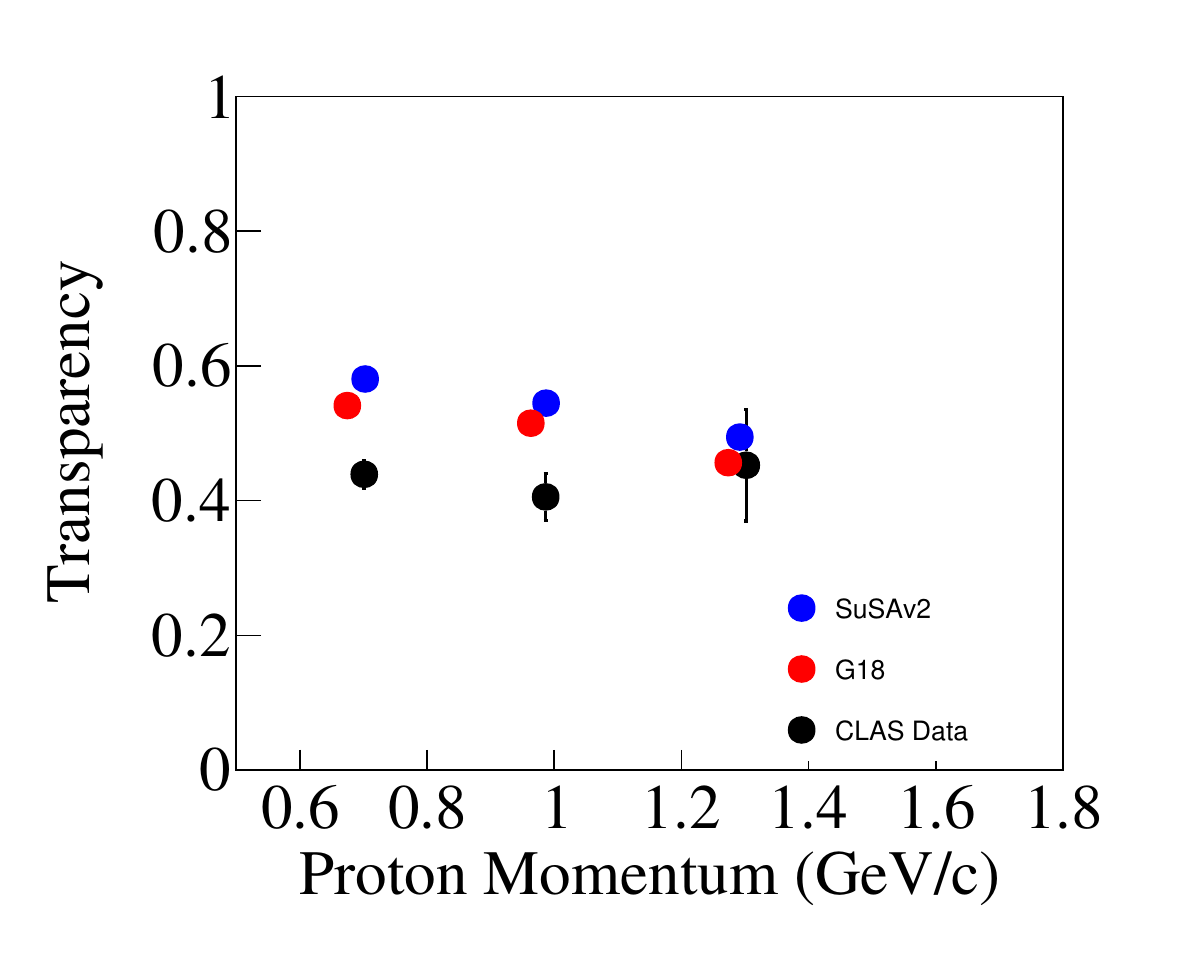}
    \caption{Proton transparency measurement for iron with CLAS data (black) compared to predictions from G18+LFG+hA2018 models (red) and SuSAv2+LFG+hN2018 models (blue).}
    \label{fig:MC-data-fe}
\end{figure}

We also compare our measured transparencies with the results of MC models. When comparing models with transparency measurements, care must be taken to match the cuts exactly~\cite{Niewczas:2019fro}. We therefore applied the same cuts to data and GENIE predictions. Figures~\ref{fig:MC-data-he}, ~\ref{fig:MC-data-c}, and \ref{fig:MC-data-fe} compare our data with GENIE simulations using SuSAv2 and G18 interaction models with different ingredients by varying FSI models and struck nucleon momentum distributions, as detailed below.

The calculations differ from the data both in shape and normalization for all targets. Data at $p_p\sim1.65\;\mathrm{GeV}/c$ have good agreement with the calculations although these data points tend to have larger uncertainties than for lower momenta. The discrepancy grows as the proton momentum decreases. At the lowest $p_p$, the difference between MC and data is largest for iron, and G18 tends to be closer to the data than SuSAv2. This is expected because both nuclear structure and FSI effects are expected to grow for larger nuclei and smaller proton momenta.
%Replacing the plots with another legen{New_Plots_for_Paper/Helium_MC_Data_newlabels.png}
%/{New_Plots_for_Paper/Carbon_MC_Data_newlabels.png}
%/{New_Plots_for_Paper/Iron_MC_Data_newlabels.png}

There are larger theoretical uncertainties for helium since the codes are often based on Fermi Gas models which often don't apply to very light nuclei. None of these models were designed for use in very light nuclei and there are no existing comparisons to helium data. The helium nucleon momentum distribution was approximated as a Fermi gas using a Fermi Momentum of $k_F=115$ MeV/c~\footnote{E. Christy, private communication}. The Fermi gas model is less accurate for lighter nuclei.  Despite this, GENIE describes the helium transparency data about the same as for the heavier nuclei.

We varied the ingredients of the GENIE calculations to see the effect of different cross section models (G18 vs. SuSAv2), different nuclear models (CFG vs. LFG) and different FSI models (hA vs. hN), as shown in Fig~\ref{fig:MC-data-c}.  Both FSI models use the same stepping method within the INC framework. Thus, both models predict the same proton transparency and differed only in the distribution of outgoing rescattered hadrons~\cite{Dytman:2021ohr}. Therefore, differences between the G18 and SuSAv2 models should be due to different QE models. The G18-based transparencies were slightly lower than the SuSAv2-based ones.  The Correlated Fermi Gas (CFG) model gave slightly lower transparencies than the Local Fermi Gas model, due to the effect of nucleon-nucleon ($NN$) correlations that reduced the number of mean-field QE $(e,e'p)$ events. Because SuSAv2 was implemented in GENIE as amplitudes, changing the underlying nucleon momentum distribution has no effect on the inclusive cross section, except for effects related to Pauli blocking. The data were corrected for the effect of correlations, but the calculation was not.
%It is important to explore the differences among the core GENIE models and more recent versions.  In Fig.~\ref{fig:MC-data-c}, single components (either the nucleon momentum distribution or the FSI model) are varied while leaving everything else the same for a carbon target.

The large discrepancy in magnitude between the base simulations and the data is surprising because the NuWro~\cite{Niewczas:2019fro} and Isaacson et al.~\cite{Isaacson:2020wlx} simulations had qualitative agreement with the previous data for carbon and iron.  Both calculations had different assumptions for both FSI, QE model, and nuclear structure.  The GENIE calculations can have problems with either nuclear structure or FSI, as shown in Sect.~\ref{sec:discussion}.

\subsection{A dependence of Transparency}
The A dependence of proton transparency is typically characterized by the power of an exponential~\cite{ONeill:1994znv,Garrow:2001di,Garino:1992ca,CLAS:2018xsb}:
\[
T(A) \propto A^\alpha.
\] 
Although this simple formula describes the data well, the parameter cannot be derived from theory. 
Previous studies found different values for $\alpha$.  O'Neill \textit{et al.}~\cite{ONeill:1994znv} found $-0.18 \le \alpha\le   -0.29$, depending on $Q^2$. Previous CLAS measurements of transparency ratios found $\alpha = -0.289\pm0.007$~\cite{CLAS:2018xsb} and $\alpha=-0.34\pm0.02$~\cite{Hen:2012fm}.  The larger value was consistent with Glauber calculations.  

In this work, we did a similar analysis for the three combined proton momentum ranges at 2.2 GeV beam energy.  Additional fits for a single proton momentum data showed little dependence on the data used. The results were an exponent of  $\alpha = -0.225\pm0.005$ for data and $\alpha = -0.17\pm0.003$ for Monte Carlo. Even though $^4$He was not well described by a Fermi gas, it still fits in with the same power law. However, inclusion of helium in our fits makes the exponent somewhat bigger.
%This likely results from the different nuclear structure of the nuclei for $A<12$.
The discrepancy between our data and the GENIE simulations is the most interesting result. All these calculations use a mean free path based on the free total nucleon-nucleon reaction cross section with modifications for $NN$ correlations by Pandharipande and Pieper~\cite{Pandharipande:1992zz} that were developed in response to the MIT-Bates data~\cite{Garino:1992ca}. 
%The discrepancy with respect to the data is likely tied up within the considerations elsewhere in this section.

\section{Discussion} \label{sec:discussion}
As shown in Refs.~\cite{Dytman:2021ohr,Pandharipande:1992zz,Frankfurt:2000ty}, transparency calculations are sensitive to both nuclear structure and FSI components of each model.  We therefore varied the ingredients of the GENIE calculations in Figs.~\ref{fig:MC-data-he}-\ref{fig:MC-data-fe} to see the effect of different cross section models (G18 vs. SuSAv2), different nuclear models (CFG vs. LFG) and different FSI models (hA vs. hN).  In this section, we discuss the role of each effect in our results.

Previous articles have studied transparency with models similar to those described above.  For example, Ref.~\cite{Niewczas:2019fro} shows NuWro results for transparency in carbon and iron.  The simulations used an LFG nuclear model and an FSI model similar to the GENIE hN2018 model. They carefully matched the conditions of each measurement and got results that are somewhat above most of the data. 

%FSI and nuclear structure effects are traditionally discussed as independent.  In reality, they are intertwined because each FSI model also requires a nuclear structure model that is ideally the same as is used for the QE vertex.

Ref.~\cite{Dytman:2021ohr} shows the effect of various QE, nuclear structure and FSI model choices that are in GENIE, NuWro, and NEUT event generators.  They found that transparency at proton momentum greater than about 500 MeV/c is largely insensitive to these choices. However, almost all models involved used the LFG nucleon momentum distribution.

\FloatBarrier
\subsection{FSI effects}
\label{se:fsi}
The FSI model will directly affect the transparency results. All models used in this analysis predict higher transparency than the data, but almost no difference was shown when changing only the FSI model. In this section, sensitivities to a variety of choices is made.

The FSI models used in this analysis are often used in accelerator-based neutrino experiments. Like most models in use, they are based on the semiclassical INC approximation in which the hadron-nucleon interaction in the nuclear medium is the free cross section with nuclear corrections.  
Most INC codes model the struck nucleon kinematics with a momentum distribution and a binding energy correction that is derived from electron scattering measurements. NuWro~\cite{Golan:2012wx} and GENIE hN2018 models are very similar.  Both have added the same nuclear medium corrections~\cite{Pandharipande:1992zz} to the hadron-nucleon interaction; this was shown to be a few percent increase to transparency~\cite{Dytman:2021ohr} and a similar decrease in $\sigma_{reac}$. 

%Event generators for neutrino experiments must provide a complete description of the final state.  Thus, relativistic optical potentials~\cite{ivanov} that are typically used in electron scattering experiments are inappropriate. Nikolopoulos et al.~\cite{nikol} show that for the simple case of QE scattering the relativistic optical potential and the INC approach give very similar results.  

The calculations presented here use FSI codes that have been largely validated against proton total reaction cross section ($\sigma_{reac}$) data~\cite{Carlson:1996ofz} rather than transparency measurements. Typically, agreement for $\sigma_{reac}$ at nucleon momenta larger than roughly 400 MeV/c can be obtained with free $pN$ interactions~\cite{Dytman:2011zz}. 

GENIE has three FSI models that can be used interchangeably due to its modular structure.    
The GENIE hA and hN models shown in Sect.~\ref{sec:results} are not the most complete models available and alternative FSI models have been added~\cite{GENIE:2021npt}. The INCL model~\cite{cugnon:2016ghr,mancusi:2014eia} has an improved nuclear treatment that is based on a mean field model with quantum corrections. It also includes a model where the recoiling protons pick up nucleons to form multinucleon clusters~\cite{Ershova:2023dbv} that decreases the proton yields. The resulting values using INCL for $\sigma_{reac}$ and transparencies are quite different than hA and hN at low proton energies but in very good agreement at the energies relevant to this measurement~\cite{Dytman:2021ohr}. 

Ref.~\cite{Dytman:2021ohr} also studied the effects of various nuclear phenomena.  Pauli blocking is important for proton momenta less than roughly 300 MeV/c.  Proton formation zones are not needed to get agreement with previous transparency data~\cite{Dutta:2012ii} and are not in use in the codes studied. Dytman et al.~\cite{Dytman:2021ohr} also showed that $NN$ correlations in NuWro~\cite{Pandharipande:1992zz} produce a significant change to transparency, but much less to $\sigma_{reac}$.
The authors suggest that $\sigma_{reac}$ is less sensitive to $NN$ correlations than transparency because FSI is much less affected by characteristics of the second nucleon than in the $eN$ vertex for QE scattering.  Hadron beams only require interaction with a single target nucleon to contribute to $\sigma_{reac}$ but the second nucleon is essential to determine whether the event is true QE or not.

Other recent articles have studied effects beyond the typical INC model.  Ershova et al.~\cite{Ershova:2023dbv} compares NuWro results with the default FSI model with a Spectral Function nuclear model to one with INCL FSI. Since the NuWro QE model is used to produce protons, this is a test that focuses on FSI. Differences in transparency are 5--10\% at the larger proton momenta. Their main conclusion notes the effects of multinucleon clusters. 
A new INC model by Isaacson et al.~\cite{Isaacson:2020wlx} studied different stepping mechanisms.  These alternative treatments vary the definition of the space around a nucleon where the interaction can occur. Both $\sigma_{reac}$ and transparency change in the same direction. Results differ by a few percent for proton momenta 0.8--1.2 GeV/c. For both studies, the effects are larger at proton energies below those studied here.

Conclusions from these studies can tentatively be made.  Both $\sigma_{reac}$ and transparency depend on the $NN$ interaction in very similar ways~\cite{Dytman:2021ohr} with the main difference coming from variation in path lengths in the residual nucleus.  Effects tend to be of opposite sign in the two observables. Therefore, any significant change to this basic ingredient that decreases the predicted transparency to get agreement with data will also increase $\sigma_{reac}$ by a similar amount. Thus, any change affecting only the $NN$ cross sections to fit our transparency results will worsen the existing agreement with $\sigma_{reac}$ data.  Some other effect would then be needed to regain agreement with the $\sigma_{reac}$ data. Nevertheless, FSI models have a number of ingredients that are not known well, e.g. the nuclear effects within FSI models and the detailed stepping strategy.  Although more studies are definitely needed, there is no obvious problem in the FSI models used here in the energy range of this measurement.

\subsection{Nuclear structure effects}

\subsubsection{Nuclear models}
Various theoretical papers discuss the role of nuclear structure effects in transparency calculations. Pandharipande and Pieper~\cite{Pandharipande:1992zz} found nuclear effects due to Pauli blocking and $NN$ correlations to be important.  They calculated both the spatial effects of correlations on the rescattering of the outgoing protons and the momentum effects on the ratio of Eq.~\ref{TAold}. The momentum effect correction was the basis for the SRC corrections used in many previous measurements~\cite{Dutta:2003yt}. The size of these corrections was questioned by Frankfurt, Strikman, and Zhalov~\cite{Frankfurt:2000ty} as discussed above. 
A notable result of Ref.~\cite{Dytman:2021ohr} is to show that the spatial effects of $NN$ correlations (i.e., the reduced probability of finding a second nucleon within a short distance of the struck nucleon) in NuWro~\cite{Pandharipande:1992zz} increase the proton transparency by roughly 10\% at all momenta. Since the effect on $\sigma_{reac}$ was much smaller, this suggests that the dominant sensitivity is in the $ep$ vertex. Isaacson et al.~\cite{Isaacson:2020wlx} used a nuclear model based on a Greens Function Monte Carlo wave function~\cite{Carlson:2014vla}. They emphasized the role of $NN$ correlations in effectively keeping nucleons apart. This effect is in both initial and final states in their simulation.

%The base GENIE models used for this measurement use the LFG momentum distribution which doesn't include the effects of $NN$ correlations.  This provides qualitative agreement with some neutrino-nucleus cross section data. The data here provide a more strict test.  More complete GENIE models became available as this work neared completion and we were able to study these aspects.

As noted in Sect.~\ref{se:fsi}, there is a 5--10\% difference between the G18 and SuSAv2 transparencies (SuSAv2 LFG hA vs. G18 LFG hA) and a similar difference between the SuSAv2 LFG and CFG model transparencies. The CFG/LFG difference verifies that the nucleon momentum distribution affects the calculated transparency values and the need for SRC corrections to nuclear structure models. Detailed agreement with data is not expected because the kinematics of the emitted nucleon are unchanged from calculations with LFG.

To more accurately include the effect of correlations, a new GENIE QE simulation for $^{12}$C based on a Correlated Basis Function Spectral Function (SF) calculation was made~\cite{Betancourt:2023uxz}.  To focus on nuclear structure, the same hN2018 FSI model as the SuSAv2 calculation was used.  This spectral function includes $NN$ correlations based on the Local Density Approximation and a depleted mean field region fit to $(e,e'p)$ data~\cite{Benhar:1989aw,Benhar:1994hw}.  The influence is best seen in the missing energy ($E_{miss}$) and missing momentum ($P_{miss}$) distributions: 
%\begin{center}
%\begin{equation}
\begin{align}
E_{\rm{miss}}&=\nu - T_{p'} - T_{A-1}\\
\vec{P}_{\rm{miss}} &= \vec{p}_{p'} - \vec{q},
\label{eqn:pmEm}
\end{align}
%\end{equation}
where $\nu$ and $\vec{q}$ are the virtual photon energy and 3-momentum, respectively, and $T_{p'}$ and $\vec{p}_{p'}$ are the kinetic energy and momentum of the scattered proton. In the SF model, most nucleons have low momentum in the ground state, often called the mean field region.  QE scattering off these nucleons populates the region $P_{\rm{miss}}<300$ MeV/c~\cite{Dutta:2003yt}.  However, about 20\% of the nucleons have $P_{\rm{miss}}>300$ MeV/c due to the effect of $NN$ correlations. This has immediate consequences for the final transparency results, as our selection cuts remove many of the $P_{\rm{miss}}>300$ MeV/c events. Since the LFG momentum distribution has no high-$P_{\rm miss}$ $NN$ correlation tail to remove, there will be an excess of events coming from the mean field region.  Therefore, simulations using it have a larger normalization than in the data. 

\begin{figure}[htb!]
    \centering
    \includegraphics[width=0.45\textwidth]{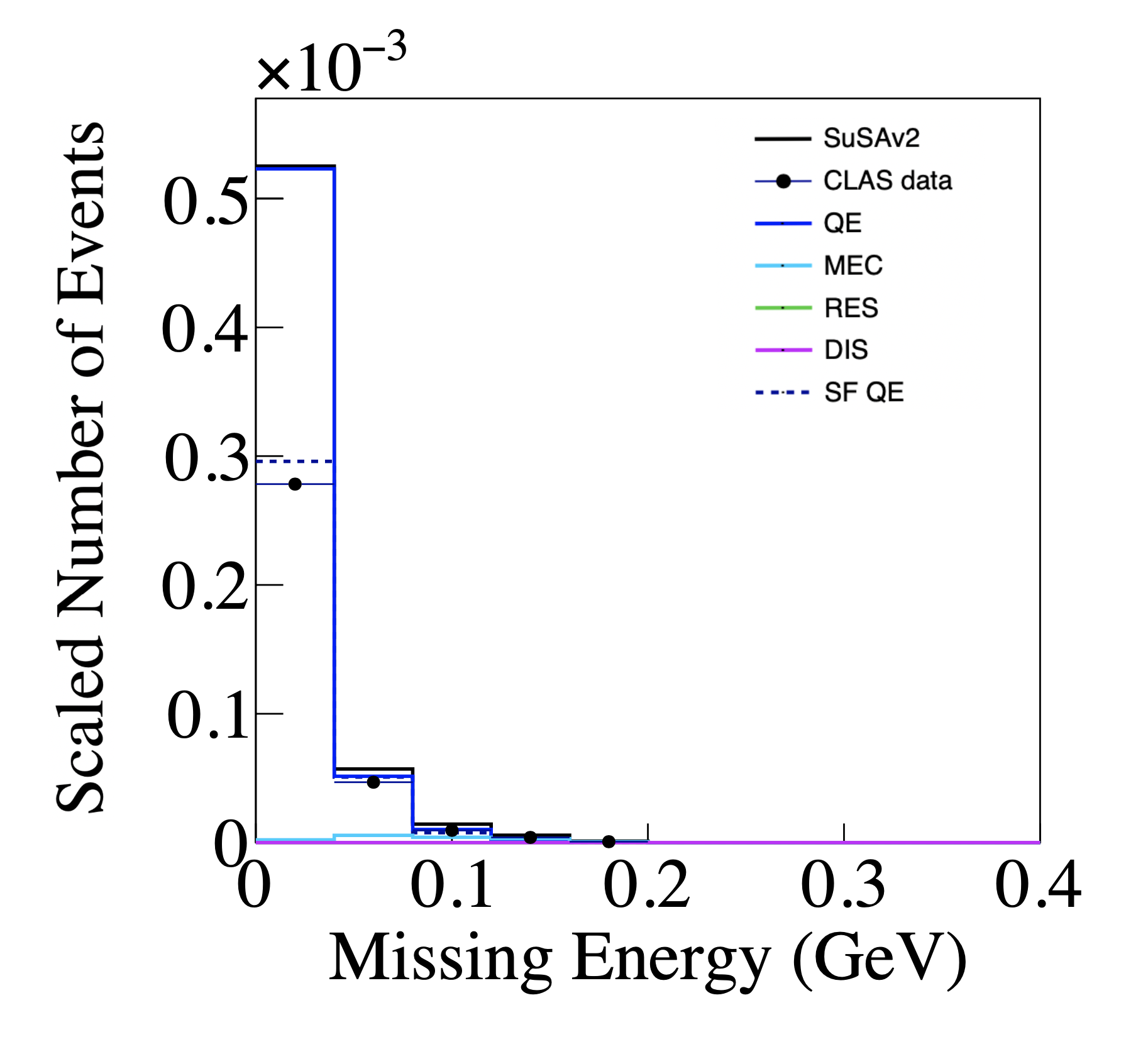}
    \includegraphics[width=0.45\textwidth]{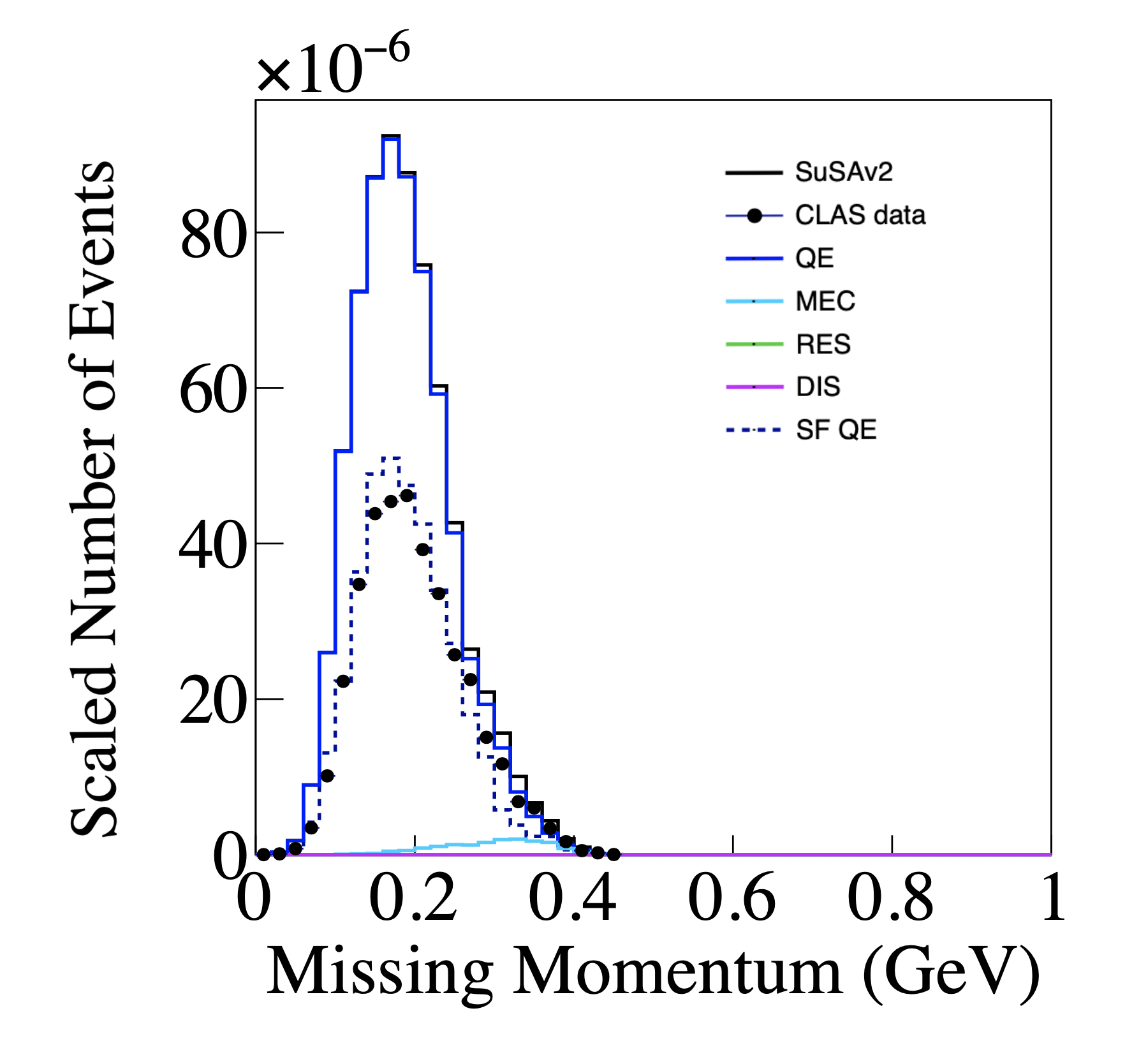}
    \caption{Number of events (luminosity normalized) as a function of reconstructed missing energy (top) and missing momentum (bottom) for 2.261 GeV C$(e,e'p)$ for $21^\circ\leq\theta_e\leq23^\circ$ after all event selection cuts. Comparison with SuSAv2 broken into different interaction components is shown (solid lines), as well as a prediction from the QE Spectral Function model (dashed line). Data has radiative corrections applied.}
    \label{fig:C12_EmPm}
\end{figure}

\begin{figure}[htb!]
    \centering
    \includegraphics[width=0.45\textwidth]{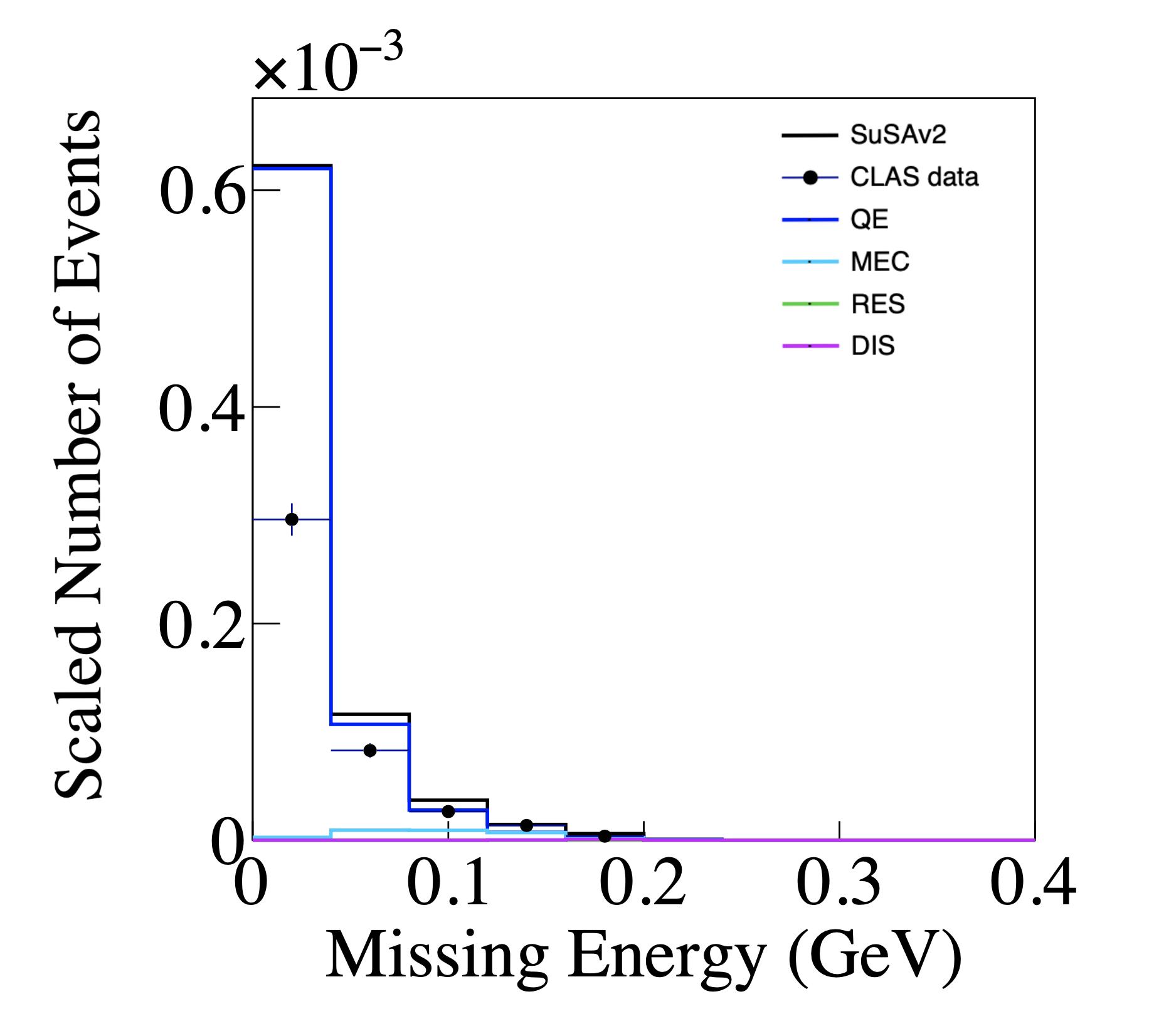}
    \includegraphics[width=0.45\textwidth]{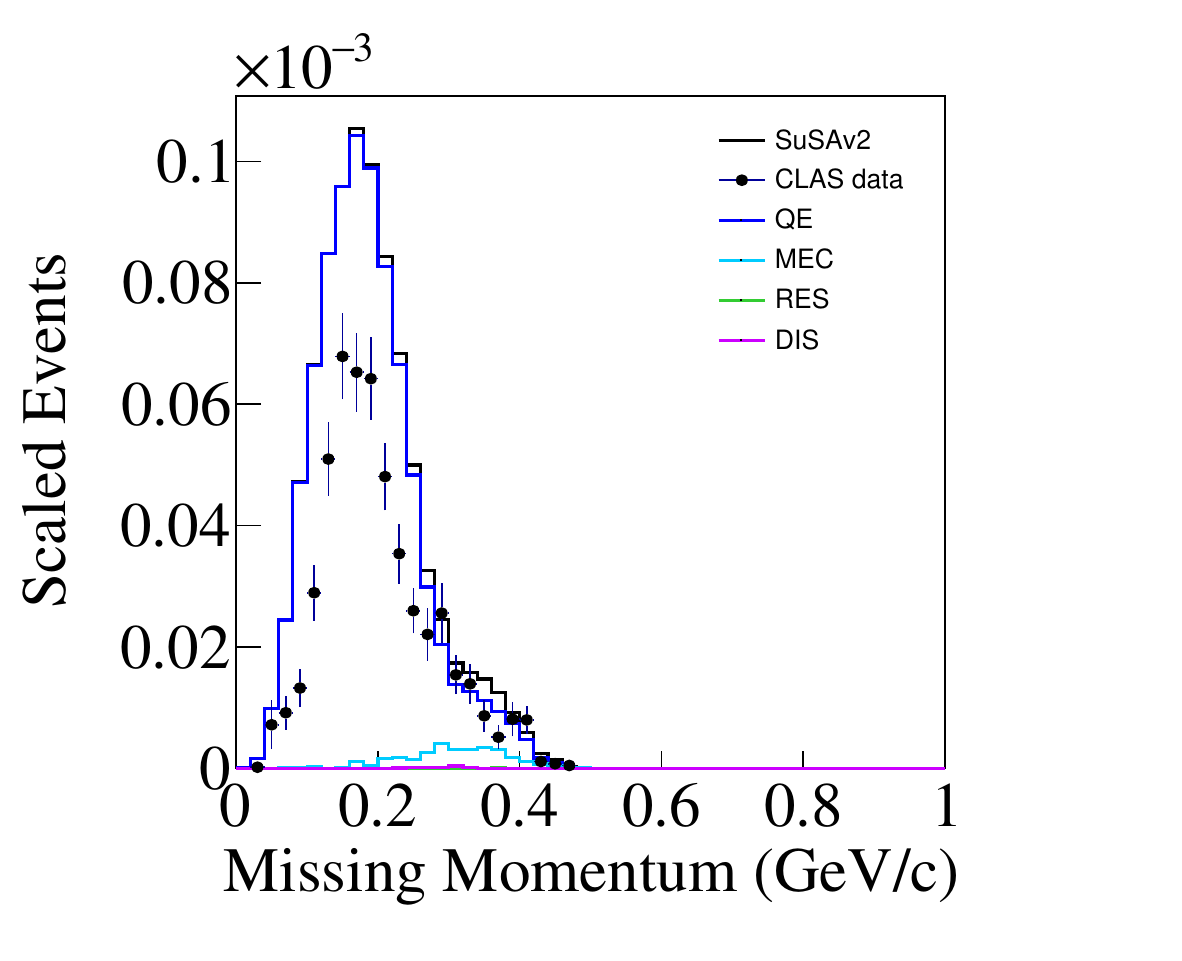}
      %Pm_56Fe_2.261000_2.png}
    \caption{Number of events (luminosity normalized) as a function of reconstructed missing energy (top) and missing momentum (bottom) for  2.261 GeV Fe$(e,e'p)$ at $28^\circ\leq\theta_e\leq31^\circ$ after all event selection cuts. Comparison with SuSAv2 broken into different interaction components is shown (solid lines). Data has radiative corrections applied.}
    \label{fig:Fe56_EmPm}
\end{figure}

Figures~\ref{fig:C12_EmPm} and~\ref{fig:Fe56_EmPm} present the $E_{miss}$ and $P_{miss}$ data distributions for the $(e,e'p)$ interaction with all experimental cuts and corrections compared to the SuSAv2 model using LFG and a separate QE calculation with the SF model. For carbon we present range 1 ($21^\circ\leq\theta_e\leq23^\circ$) and iron range 2 ($28^\circ\leq\theta_e\leq31^\circ$).  We ignore the small kinetic energy of the recoiling nucleus, $T_{A-1}$.  The SuSAv2 model overpredicts both the carbon and iron data while the carbon SF calculation describes the shape and magnitude of the $E_{miss}$ and $P_{miss}$ distributions remarkably well.  While we do not have SF calculations for iron (Fig.~\ref{fig:Fe56_EmPm}), measurements of the spectroscopic factors in $(e,e'p)$ experiments on nuclei with $7 < A < 208$ show that the depletion of the low-momentum (mean field) region is essentially A-independent~\cite{Lapikas:1993uwd}. We therefore conclude that a large portion of the discrepancy between GENIE results and the data come from mis-modeling of the nuclear ground state in the calculations.

\section{Conclusion} \label{sec:conclusions}
Interest in hadron transparency is increasing  because it can be a better technique than hadron-nucleus scattering to validate FSI models in MC simulations of neutrino experiments~\cite{Dytman:2021ohr,Isaacson:2020wlx,Niewczas:2019fro}. This is important because FSI is an important component of the estimated systematic uncertainties for both neutrino oscillation and neutrino cross section measurements. 

To that end, we analyzed the CLAS e2a dataset to extract the nuclear transparency using a new method.  We selected narrow bins in $\theta_e$ to examine bins of proton momentum from 0.8 to 1.65 GeV/c for $^4$He, carbon, and iron targets. The large acceptance of CLAS was important to establish the validity of the measurement. 

This analysis used true QE events to sample recoil protons whose kinematics can be anticipated based on the kinematics of $(e,e’p)$ scattering. The range of momenta considered overlaps that of previous results from Jefferson Lab~\cite{Garrow:2001di,Rohe:2005vc}, SLAC~\cite{ONeill:1994znv}, and MIT-Bates~\cite{Garino:1992ca}. 
%The set of systematic uncertainties considered for this analysis is broader than previous measurements. Nevertheless, error bars are of comparable size except at the highest momentum transfers where the statistics were poor.  
The present analysis largely confirms previous results for carbon and iron, except for possible discrepancies at the lowest proton momenta. The data for helium are new and interesting because its nuclear structure is treated differently in theoretical models than for the heavier nuclei.

In this work, we used a different method to extract transparency than in previous measurements~\cite{Dutta:2003yt,ONeill:1994znv}. We defined the transparency as the ratio of $(e,e'p)$ quasielastic events with a proton to $(e,e')$ QE events that should have a detected proton.  We reduced the effects of non-QE reactions by requiring that the $(e,e')$ events were from the low-energy-loss side of the QE peak and then we corrected for the small fraction of non-QE (e.g., MEC or $2p2h$) events remaining.  Most previous measurements defined transparency as the ratio of $(e,e'p)$ quasielastic events to the expected number of calculated PWIA events.  The two methods differ in their model dependence.  However, both methods needed to be corrected for protons knocked out from  SRC pairs.  It is interesting to note that the previous measurements required an enhancement of the denominator because the calculation is incomplete and the new method requires removal of events that are not true QE.  The previous measurements used an SRC correction with no $Q^2$ dependence that has been called into question~\cite{Frankfurt:2000ty}. The MIT-Bates measurement did not correct for $NN$ correlations.
Previous measurements did not explicitly correct for the $2p2h$ background because it was expected to be small and calculations such as were employed in this analysis were not yet available. A full understanding of the $2p2h$ contamination in the signal region is an important ingredient in our analysis to obtain a true QE sample. 
 
Although both MC calculations agree with the transparency data at high proton momentum, the deviation between data and calculations grows as the proton momentum decreases so that the calculations are well above the data for all three targets at the lowest momenta.
Although poorer agreement with helium is expected due to the Fermi Gas approximation for the momentum distribution, discrepancies seen for helium are similar to those found for heavier nuclei. The MC-data discrepancy for carbon is examined in some detail in Sect.~\ref{sec:discussion}. 

Problems with MC in describing transparency can come from either nuclear structure or FSI treatments. Although transparency calculations with various FSI models qualitatively agree with previous data~\cite{Niewczas:2019fro}, significant dependence on nuclear structure is noted~\cite{Dytman:2021ohr,Isaacson:2020wlx}. 
The MC calculations available for this measurement used an LFG nucleon momentum distribution; these are commonly used in comparisons with neutrino cross section data~\cite{MicroBooNE:2023cmw,MINERvA:2018hba}.  
Further calculations that examine nuclear structure were made for the carbon target. The Spectral Function model~\cite{Betancourt:2023uxz} has about 20\% of the nucleons populating a high-$P_{\rm miss}$ tail due to $NN$ correlations and a different model for the QE interaction inside the nucleus. The Spectral Function model calculation describes $P_{\rm miss}$ and $E_{\rm miss}$ distributions much better than the LFG model. SF calculations of transparency were not attempted because they are only available for the QE process.
 
Although two FSI models were used for comparison, this does not cover all the possibilities.  A review of relevant literature on FSI in Sect.~\ref{sec:discussion} shows that nuclear corrections in FSI contributions to transparency can be significant. Studies show that Pauli blocking, nucleon pickup effects and most nuclear effects are largely important at proton momenta below those studied in this article.  However, other effects such as the stepping mechanism and SRC correlations have been shown to be important at higher momenta. Although the stepping mechanism affects both $\sigma_{reac}$ (the proton total reaction cross section) and transparency, SRC correlations have a larger effect on transparency.
It should also be noted that both $\sigma_{reac}$ and transparency are primarily sensitive to the initial interaction of the recoiling proton.  Therefore, transparency is not sensitive to the full FSI model.

To decrease the calculated transparencies and bring them into agreement with the data would require significantly increasing the nucleon-nucleon scattering cross section, likely causing a conflict with proton-nucleus scattering measurements~\cite{Dytman:2021ohr}.  However, it is possible that other effects such as those discussed in Sect.~\ref{sec:discussion} could also resolve the discrepancy. More studies are needed to establish the importance of these effects. 

Thus, while the SuSAv2 and G18 calculations agree well with inclusive data~\cite{SuSAv2QE,e4nu-egenie:2020tbf}, there are discrepancies in the  transparency calculations and in comparisons to $(e,e'p)$ data. Spectral function calculations for quasielastic electron scattering agree much better with the $(e,e'p)$ data.  Evaluations of the SF nuclear model with neutrino cross sections are more complicated than what was shown here. Studies with the SF model show mixed agreement with neutrino cross section data~\cite{MINERvA:2018hba,MINERvA:2019ope,T2K:2023qjb}. 

The surprising conclusion of this work is that nuclear structure in event generator codes can significantly affect transparency calculations. We look forward to improved calculations including modern information on nuclear structure (e.g., spectral function calculations) that can better test FSI models in proton transparency.  Although less detailed studies of FSI were made, no obvious problems were observed.  A comparison of $\sigma_{reac}$ with transparency is very profitable as they have different dependencies on nuclear effects. The transparency measurement presented here provides a new and important test of these models.

\section{Acknowledgments}
The authors acknowledge the efforts of the staff of the Accelerator
and Physics Divisions at Jefferson Lab that made this experiment
possible. The authors thank Luke Pickering for useful discussions. The
analysis presented here was carried out as part of the Jefferson Lab
Hall B Data-Mining project supported by the U.S. Department of Energy
(DOE). This manuscript has been authored in part by FermiForward
Discovery Group, LLC under Contract No. 89243024CSC000002 with the
U.S. Department of Energy, Office of Science, Office of High Energy
Physics.  The research was supported also by DOE, the National Science
Foundation, the Israel Science Foundation, the Chilean Agencia
Nacional de Investigacion y Desarrollo (ANID), the French Centre
National de la Recherche Scientifique and Commissariat a l’Energie
Atomique, the French-American Cultural Exchange, the Italian Istituto
Nazionale di Fisica Nucleare, the Deutsche Forschungsgemeinschaft
(DFG), the National Research Foundation of Korea, and the UK’s Science
and Technology Facilities Council. PC acknowledges support from
project PROMTEO/2019/083. This work was supported in part by the
U.S. Department of Energy, Office of Science, Office of Workforce
Development for Teachers and Scientists (WDTS) under the Science
Undergraduate Laboratory Internships Program (SULI).  This work has been supported by Laboratory
Directed Research and Development (LDRD) funding from Fermilab and Department of Energy. This project has been supported by the European Union Horizon 2020 research and innovation program under the Marie Sklodowska-Curie grant agreement No. 674896 (Elusives, H2020-MSCA- ITN- 2015-674896). Jefferson Science Associates operates the Thomas Jefferson National Accelerator Facility for the DOE, Office of Science, Office of Nuclear Physics under contract DE-AC05-06OR23177. The raw data from this experiment are archived in Jefferson Lab’s mass storage silo.

\section{Appendix}

The CLAS spectrometer has six essentially identical detectors arranged at different azimuthal angles around the beam line. Therefore, the only unpolarized cross-section differences among the sectors should come from sector-specific inefficient or bad detector elements. Here, we supply information broken down by sector. %In addition, we examine the cross section results for the numerator and denominator.

\begin{figure}[htb!]
    \centering
    {\includegraphics[width=0.48\textwidth]{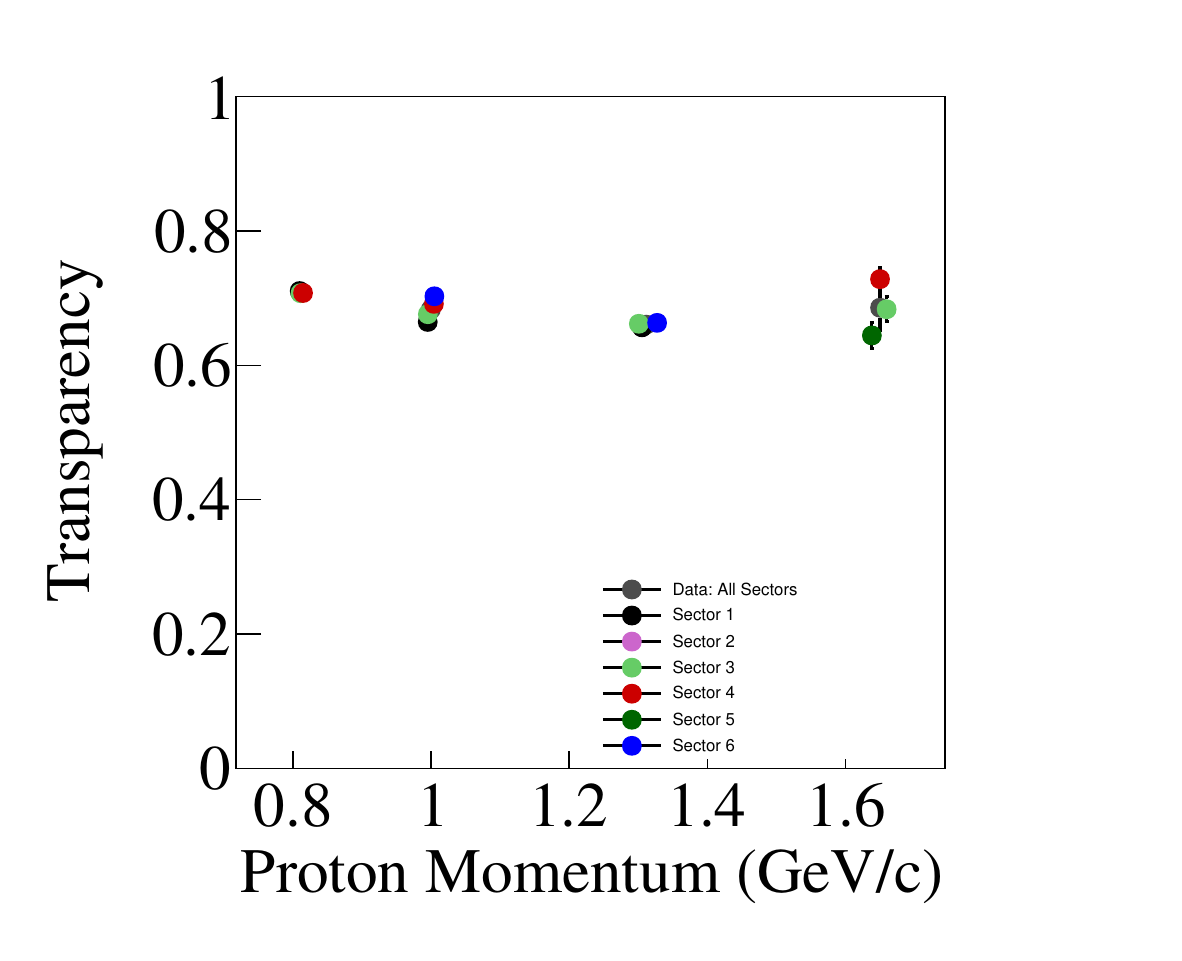}}
    \caption{Transparency of Helium broken down into sectors used in the analysis.  For each proton momentum, proton sectors according to Tab.~\ref{tab:cuts-he-2.2} are used in the analysis.}
      \label{transparencyHepersector}
\end{figure}

\begin{figure}[htb!]
    \centering
    {\includegraphics[width=0.48\textwidth]{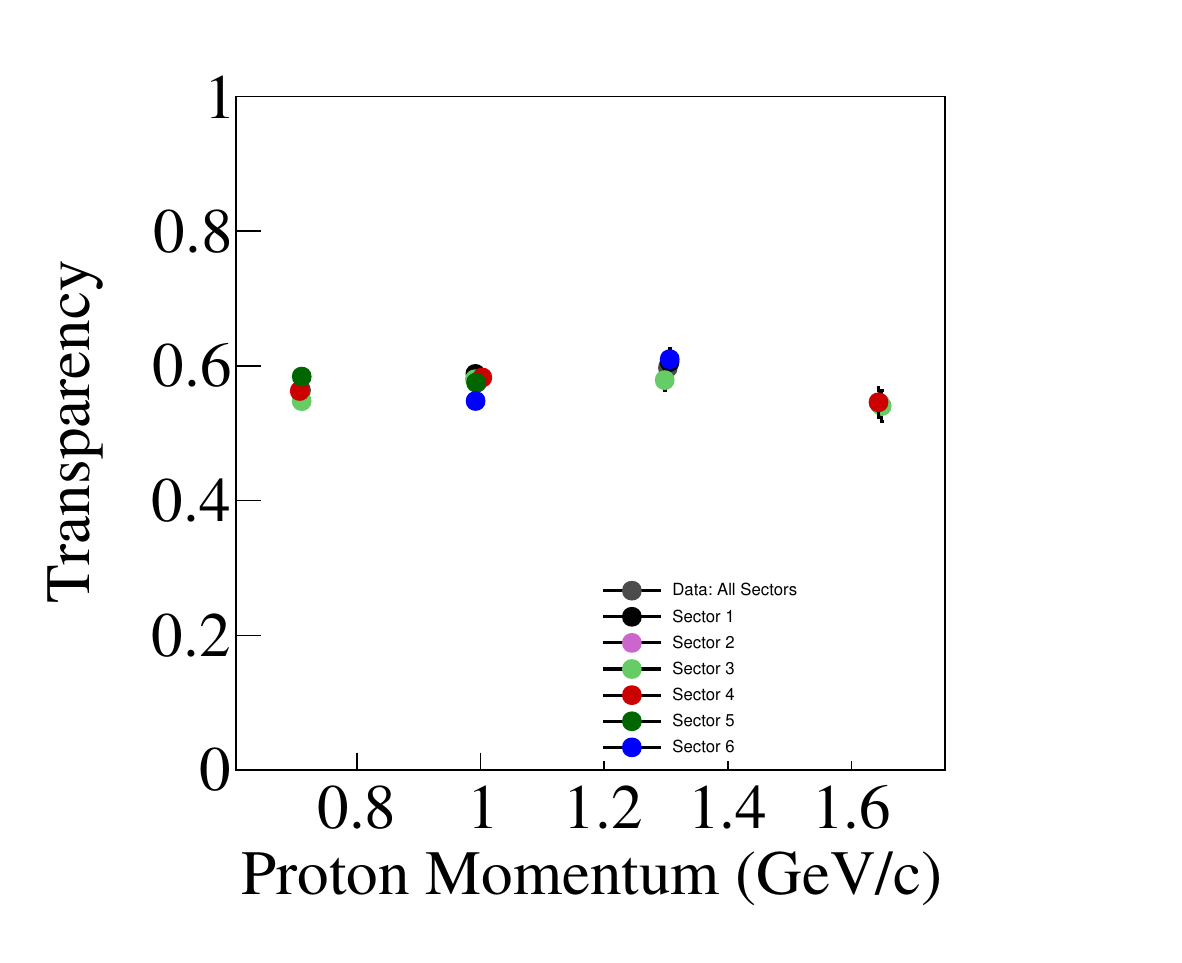}} 
    {\includegraphics[width=0.48\textwidth]{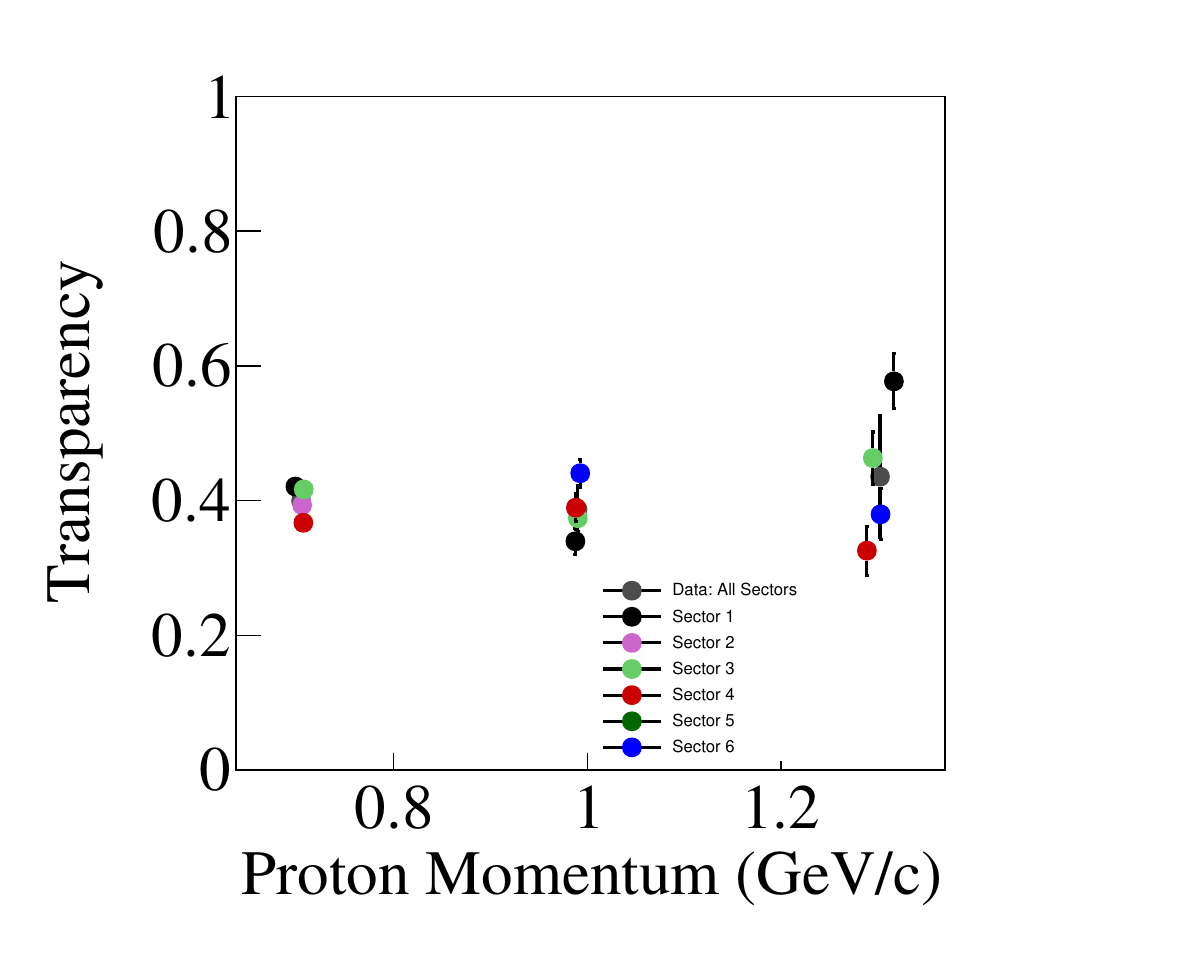}}
    \caption{Transparency of carbon (top) and iron (bottom) broken down into sectors used in the analysis.  For each proton momentum, sectors according to Tab.~\ref{tab:cuts-c-2.2} (carbon) and Tab.\ref{tab:cuts-fe-2.2} (iron) are used in the analysis.}
     \label{transparencyCFepersector}
\end{figure}

Figures \ref{transparencyHepersector} and \ref{transparencyCFepersector}  show the transparency for each sector used in the analysis for helium, carbon and iron. The variation among the sectors is reasonable. The uncertainties shown are purely statistical and the sector-to-sector variation is larger than expected from the statistical uncertainties. We leverage this to assign a systematic uncertainty from the variance in the measured transparencies in each sector (see Sect.~\ref{sec:systematics}).


\bibliographystyle{apsrev}
\bibliography{references}

\end{document}